\documentclass[lettersize,journal]{IEEEtran}
\usepackage{amssymb}
\usepackage{algorithm}
\usepackage{algorithmic}
\usepackage{amsmath,amsfonts}
\usepackage{amsthm}
\usepackage{array}
\usepackage{arydshln}
\usepackage{bbding}
\usepackage{bigints}
\usepackage{booktabs}
\usepackage[caption=false,font=normalsize,labelfont=sf,textfont=sf]{subfig}
\usepackage{cite}
\usepackage{color}
\usepackage{diagbox}
\usepackage{epsfig,latexsym}
\usepackage{epstopdf}
\usepackage{graphicx}
\usepackage{fancyhdr}
\usepackage{float}
\usepackage{flushend}
\usepackage{indentfirst}
\usepackage{lastpage}
\usepackage{makecell}
\usepackage{mathtools}
\usepackage{multirow}
\usepackage{pifont}
\usepackage{psfrag}
\usepackage{setspace}
\usepackage{stfloats}
\usepackage{subfig}
\usepackage{subfloat}
\usepackage{textcomp}
\usepackage{times}
\usepackage{verbatim}
\usepackage{url}
\usepackage{hyperref}
\usepackage{balance}
\usepackage{cleveref}
\newcommand{\cmark}{\ding{52}}

\theoremstyle{remark}

\allowdisplaybreaks[4]

\begin{document}
\title{Stacked Intelligent Metasurfaces for Holographic MIMO Aided Cell-Free Networks}
\author{Qingchao Li, \textit{Graduate Student Member, IEEE},
Mohammed El-Hajjar, \textit{Senior Member, IEEE},\\
Chao Xu, \textit{Senior Member, IEEE},
Jiancheng An, \textit{Member, IEEE},
Chau Yuen, \textit{Fellow, IEEE},\\
and Lajos Hanzo, \textit{Life Fellow, IEEE}

\thanks{The work of Chau Yuen was supported by the Ministry of Education, Singapore, under its Ministry of Education (MOE) Tier 2 (Award number MOE-T2EP50220-0019). The work of Lajos Hanzo was supported by the Engineering and Physical Sciences Research Council projects EP/W016605/1, EP/X01228X/1, EP/Y026721/1 and EP/W032635/1 as well as of the European Research Council's Advanced Fellow Grant QuantCom (Grant No. 789028). \textit{(Corresponding author: Lajos Hanzo.)}

Qingchao Li, Mohammed El-Hajjar, Chao Xu and Lajos Hanzo are with the School of Electronics and Computer Science, University of Southampton, Southampton SO17 1BJ, U.K. (e-mail: qingchao.li@soton.ac.uk; meh@ecs.soton.ac.uk; cx1g08@ecs.soton.ac.uk; lh@ecs.soton.ac.uk).

Jiancheng An and Chau Yuen are with the School of Electrical and Electronics Engineering, Nanyang Technological University, Singapore 639798 (e-mail: jiancheng.an@ntu.edu.sg; chau.yuen@ntu.edu.sg).}}

\maketitle

\begin{abstract}
Large-scale multiple-input and multiple-output (MIMO) systems are capable of achieving high date rate. However, given the high hardware cost and excessive power consumption of massive MIMO systems, as a remedy, intelligent metasurfaces have been designed for efficient holographic MIMO (HMIMO) systems. In this paper, we propose a HMIMO architecture based on stacked intelligent metasurfaces (SIM) for the uplink of cell-free systems, where the SIM is employed at the access points (APs) for improving the spectral- and energy-efficiency. Specifically, we conceive distributed beamforming for SIM-assisted cell-free networks, where both the SIM coefficients and the local receiver combiner vectors of each AP are optimized based on the local channel state information (CSI) for the local detection of each user equipment (UE) information. Afterward, the central processing unit (CPU) fuses the local detections gleaned from all APs to detect the aggregate multi-user signal. Specifically, to design the SIM coefficients and the combining vectors of the APs, a low-complexity layer-by-layer iterative optimization algorithm is proposed for maximizing the equivalent gain of the channel spanning from the UEs to the APs. At the CPU, the weight vector used for combining the local detections from all APs is designed based on the minimum mean square error (MMSE) criterion, where the hardware impairments (HWIs) are also taken into consideration based on their statistics. The simulation results show that the SIM-based HMIMO outperforms the conventional single-layer HMIMO in terms of the achievable rate. We demonstrate that both the HWI of the radio frequency (RF) chains at the APs and the UEs limit the achievable rate in the high signal-to-noise-ratio (SNR) region.
\end{abstract}

\section{Introduction}
\IEEEPARstart{I}{n} the fifth generation (5G) wireless systems, large-scale multiple-input and multiple-output (MIMO) systems have been harnessed for providing significantly increased throughput by employing a large number of antennas at the base station (BS)~\cite{wang2020joint}, \cite{dai2022delay}, \cite{du2022tensor}, \cite{li2023uav}, \cite{gao2023deep}. However, they require numerous active radio frequency (RF) chains, which results in excessive hardware cost and energy consumption. Hence, some authors have focused their attention on the design of low-cost and energy-efficient solutions~\cite{wang2018finite}, \cite{yang2020secrecy}, \cite{yang2020performance}, \cite{nguyen2021linear}, \cite{li2022reconfigurable_tvt}. As an attractive design alternative, holographic MIMO (HMIMO) solutions rely on an intelligent software reconfigurable paradigm in support of improved hardware efficiency and energy efficiency. They achieve this ambitious objective by utilizing a spatially near-continuous aperture and holographic radios having reduced power consumption and fabrication cost~\cite{huang2020holographic}, \cite{li2022reconfigurable_iot}, \cite{li2023reconfigurable}, \cite{yoo2023sub}, \cite{deng2023reconfigurable}, \cite{gong2023holographic}. The recent progress in channel modeling and efficient channel estimation conceived for HMIMO systems is reported in~\cite{an2023tutorial}.

Currently, the hardware architectures of HMIMO are mainly based on reconfigurable refractive surfaces (RRS)~\cite{zeng2022reconfigurable}, reconfigurable holographic surfaces (RHS)~\cite{deng2021reconfigurable_tvt}, \cite{deng2022hdma}, \cite{deng2022reconfigurable_wc}, \cite{hu2022holographic}, \cite{deng2022holographic}, \cite{hu2023holographic}, \cite{wei2022multi}, \cite{wu2023two}, and dynamic metasurface antennas (DMA)~\cite{shlezinger2019dynamic}, \cite{wang2019dynamic}, \cite{you2022energy}, \cite{li2023near}.

\subsubsection{Reconfigurable refractive surfaces}
It is infeasible to realize HMIMO schemes relying on a large number of conventional RF chains and active antennas due to the excessive power consumption~\cite{gong2023holographic}. As a remedy, Zeng \textit{et al.}~\cite{zeng2022reconfigurable} employed a RRS illuminated by a single RF chain at the BS for creating an energy-efficient HMIMO. A substantial beamforming gain can be achieved by adjusting the coefficient of each RRS element. Both their theoretical analysis and simulation results show that the RRS-aided HMIMO has higher energy efficiency than the conventional MIMO systems using phased arrays.

\subsubsection{Reconfigurable holographic surfaces}
In~\cite{deng2021reconfigurable_tvt}, \cite{deng2022hdma}, \cite{deng2022reconfigurable_wc}, Deng \textit{et al.} proposed a RHS based architecture. Their digital beamformer relying on the state-of-the-art (SoA) zero-forcing (ZF) transmit precoding method and a holographic beamformer are jointly optimized at the BS and the RHS, respectively. The simulation results showed that the RHS-based hybrid beamformer achieves a higher sum-rate than the SoA massive MIMO based hybrid beamformer relying on phase shift arrays. To reduce the configuration complexity, Hu \textit{et al.}~\cite{hu2022holographic} proposed a holographic beamforming scheme based on amplitude-controlled RHS elements having limited resolution. They showed that the holographic beamformer associated with 2-bit quantized amplitude resolution achieves a similar sum-rate as that associated with continuous amplitude values. Furthermore, in~\cite{deng2022holographic} the RHS beamformer is employed for ultra-dense low-Earth-orbit (LEO) satellite communications to compensate for the severe path-loss of satellite communications. The simulation results showed that the RHS provides a more cost-effective solution for pursuing higher data rate than the phased array architecture. The impact of the number of radiation elements on the sum-rate of RHS-based LEO satellite communications is further investigated in~\cite{hu2023holographic}. Explicitly, the authors theoretically analyzed the minimum number of RHS elements required for the sum-rate of the RHS-aided system to exceed that of the phased array system. Furthermore, Wei \textit{et al.}~\cite{wei2022multi} proposed a low-complexity ZF beamformer based on utilizing Neumann series expansion to replace the matrix inversion operation in multi-user RHS-based MIMO communications. To reduce the channel state information (CSI) estimation overhead, Wu \textit{et al.}~\cite{wu2023two} minimized the transmit power of the RHS-based holographic MIMO based on a two-time scale beamformer. Specifically, this holographic beamformer was designed based on the statistical CSI, and then the instantaneous CSI of the equivalent channel links was estimated and utilized for designing the digital transmit precoding matrix.

\subsubsection{Dynamic metasurface antennas}
The dynamic metasurface antenna consists of multiple microstrips and each microstrip is composed of a multitude of sub-wavelength and frequency-selective resonant metamaterial radiating elements, which can be employed to realize low-cost, power-efficient and size compact antenna arrays~\cite{shlezinger2019dynamic}, \cite{wang2019dynamic}, \cite{you2022energy}, \cite{li2023near}. In the DMA, the information is beamformed by linearly combining the radiation observation from all metamaterial elements in each microstrip. The mathematical model of DMA-based massive MIMO systems were firstly proposed by Shlezinger \textit{et al.} in~\cite{shlezinger2019dynamic}, where the fundamental limits of DMA-aided uplink communications was also investigated. To approach these limits, the alternating optimization algorithm was employed for designing practical DMAs for arbitrary multipath channels and frequency selectivity profiles. Furthermore, the achievable sum-rate of DMA-based downlink massive MIMO systems was characterized in~\cite{wang2019dynamic}, and an efficient alternating algorithm was proposed for dynamically configuring the DMA weights to maximize the achievable sum-rate. It was shown that the fundamental limits of DMA-based massive MIMO systems are comparable to those of conventional MIMO systems based on ideal antenna arrays. In~\cite{you2022energy}, You \textit{et al.} employed DMA to realize large-scale antenna arrays having reduced physical size, hardware cost, and power consumption. Specifically, the energy efficiency of the DMA-based massive MIMO system was maximized by the Dinkelbach transform, alternating optimization, and deterministic equivalent methods. The simulation results showed that higher energy efficiency can be achieved by the DMA-based massive MIMO architecture than by the conventional fully-digital and hybrid massive MIMO systems. Furthermore, Li \textit{et al.}~\cite{li2023near} proposed the power-efficient DMA, which operate at high frequencies and realize extremely large-scale MIMO (XL-MIMO) schemes. The DMA-based XL-MIMO architecture is composed of the conventional digital beamformer and the DMA-based holographic beamformer. Specifically, in the holographic beamformer, the DMAs can be configured based on three different modes, including continuous-amplitude configurations, binary-amplitude configurations and Lorentzian-constrained phase configurations~\cite{li2023near}. An efficient successive convex approximation based alternating direction method of multipliers (ADMM) aided algorithm was proposed for optimizing the digital beamformer and the DMA-based holographic beamformer in an alternating manner. The associated simulation results showed that the DMA-based array has lower hardware overhead and power consumption than the conventional hybrid massive MIMO beamformer.

The above holographic MIMO architectures are based on a single-layer metasurface. To further improve both the spatial-domain gain and the beamformer's degree-of-freedom, An \textit{et al.}~\cite{an2023stacked} proposed a HMIMO system based on stacked intelligent metasurfaces (SIM), which is composed of stacked multi-layer reconfigurable surfaces to carry out advanced signal processing directly in the native electromagnetic (EM) wave regime without digital beamformer. The gradient descent algorithm is employed for optimizing the phase shifts of the elements in all layers of the metasurfaces to maximize the sum-rate. The simulation results show that the SIM architecture outperforms its single-layer metasurface counterparts. The wave-based beamforming relying on the SIM is benefit of simplifying hardware architecture and improving computational efficiency. Furthermore, an SIM has the application of performing interference cancellation for multiple access and enabling integrated sensing and communications~\cite{an2024stacked_earlyaccess}.

However, the above contributions assume idealized perfect RF hardware both at the BSs and the UEs, which is impractical. Furthermore, the above HMIMO architectures are tailored for cellular networks, where the cell-edge users (UEs) suffer from low data-rate both due to the increased BS-UE distance and owing to the inter-cell interference. As a design alternative, the cell-free concept mitigates the signal path loss and the inter-cell interference by deploying a set of distributed access points (APs) for cooperatively serving UEs without cell boundaries~\cite{chen2020structured}. Specifically, the ZF method and the minimum mean square error (MMSE) method can be employed for designing the beamformers in the centralized algorithm. To reduce the required overhead of CSI-sharing between APs, a distributed algorithm can be employed based on the maximum ratio transmission (MRT) or the maximum ratio combining (MRC) criteria, albeit while at the cost of a performance degradation. Furthermore, in the distributed optimization algorithm of the cell-free system, the cooperation between APs is promising in terms of harnessing parallel computing resources and achieving almost the same data rate as the centralized algorithm~\cite{xu2023algorithm}. To deal with these challenges, in this paper we propose an SIM assisted HMIMO architecture for cell-free networks, while directly considering the signal distortion resulting from the realistic hardware impairments (HWIs) of the RF chains both at the APs and the UEs. Against this background, our contributions are detailed as follows, while Table~\ref{Table_literature} explicitly contrasts them to the literature at a glance.

\begin{table}[!t]
\footnotesize
\begin{center}
\caption{Contrasting the novelty of our paper to the existing MIMO techniques in the literature~\cite{zeng2022reconfigurable}, \cite{deng2021reconfigurable_tvt}, \cite{deng2022hdma}, \cite{deng2022reconfigurable_wc}, \cite{hu2022holographic}, \cite{deng2022holographic}, \cite{hu2023holographic}, \cite{wei2022multi}, \cite{wu2023two}, \cite{shlezinger2019dynamic}, \cite{wang2019dynamic}, \cite{you2022energy}, \cite{li2023near}, \cite{an2023stacked}, \cite{an2024stacked_earlyaccess}, \cite{chen2020structured}, \cite{xu2023algorithm}.}
\label{Table_literature}
\begin{tabular}{|p{33pt}|p{39pt}|p{34pt}|p{29pt}|p{40pt}|}
\hline
     & Metasurface technique & Multi-user system & Cell-free network & Transceiver hardware impairments \\
\hline
    Our paper & SIM & \makecell[c]{\cmark} & \makecell[c]{\cmark} & \makecell[c]{\cmark} \\
\hdashline
    \cite{zeng2022reconfigurable} & RRS &   &   &   \\
\hdashline
    \cite{deng2021reconfigurable_tvt,deng2022hdma,deng2022reconfigurable_wc,
    hu2022holographic,deng2022holographic,hu2023holographic}  & RHS & \makecell[c]{\cmark} & & \\
\hdashline
    \cite{wei2022multi}  & RHS & \makecell[c]{\cmark} & & \\
\hdashline
    \cite{wu2023two} & RHS & \makecell[c]{\cmark} & & \\
\hdashline
    \cite{shlezinger2019dynamic} & DMA & \makecell[c]{\cmark} & & \\
\hdashline
    \cite{wang2019dynamic} & DMA & \makecell[c]{\cmark} & & \\
\hdashline
    \cite{you2022energy} & DMA & \makecell[c]{\cmark} & & \\
\hdashline
    \cite{li2023near} & DMA & \makecell[c]{\cmark} & & \\
\hdashline
    \cite{an2023stacked,an2024stacked_earlyaccess} & SIM & \makecell[c]{\cmark} & & \\
\hdashline
    \cite{chen2020structured} & Conventional antennas & \makecell[c]{\cmark} & \makecell[c]{\cmark} & \\
\hdashline
    \cite{xu2023algorithm} & Conventional antennas & \makecell[c]{\cmark} & \makecell[c]{\cmark} & \\
\hline
\end{tabular}
\end{center}
\end{table}

\begin{itemize}
  \item We conceive an SIM-based HMIMO architecture for the uplink of a cell-free network, where the SIM is employed at the APs to attain high spectral- and energy-efficient information transfer. The distributed operation is employed for the SIM-based cell-free network. Specifically, at each AP the hybrid beamformer coefficients and the receiver combining (RC) vectors are jointly optimized to acquire a local estimate of the information arriving from the UEs. Afterwards, the central processing unit (CPU) uses the data detected by all APs to carry out the final detection of each UE's data.
  \item Since designing the hybrid beamformer coefficient of the SIM coefficients and the RC vectors at each AP is a non-convex problem, we propose a low-complexity layer-by-layer iterative optimization algorithm. Specifically, when the RC vectors are given, the coefficients of the intelligent metasurface are optimized on a layer-by-layer basis. By contrast, when the coefficients of all layers of the intelligent metasurface are given, the RC vectors are designed based on the MRC criterion. The SIM coefficients and the RC vectors of each AP are alternately optimized until reaching convergence.
  \item For recovering the information gleaned from the UEs, the weight vector of the CPU used for combining the local detections arriving from all APs is designed based on the MMSE criterion harnessed for maximizing the signal-to-noise-ratio (SNR) of the received signal. Furthermore, since having RF hardware impairments at APs and UEs is inevitable, we take them into account in the weight vector design by exploiting their statistics.
  \item Our numerical results show that the average achievable rate of our SIM-based HMIMO architecture in the cell-free network outperforms the conventional single-layer intelligent surface aided HMIMO. Furthermore, owing to the HWIs at APs and UEs, the achievable rate saturates at high SNRs without reaching its theoretical maximum.
\end{itemize}

The rest of this paper is organized as follows. In Section~\ref{System_Model}, we present the system model, while the beamformer design is described in Section~\ref{Beamforming_Design}. Our simulation results are presented in Section~\ref{Numerical_and_Simulation_Results}, while we conclude in Section~\ref{Conclusion}.

\textit{Notations:} Vectors and matrices are denoted by boldface lower and upper case letters, respectively, $(\cdot)^{\text{T}}$, $(\cdot)^{\dag}$ and $(\cdot)^{\text{H}}$ represent the operation of transpose, conjugate and Hermitian transpose, respectively, $\odot$ represents the Hadamard product operation, $|a|$ and $\angle a$ denote the amplitude and angle of the complex scalar $a$, respectively, $\|\mathbf{a}\|$ denotes the norm of the vector $\mathbf{a}$, $\mathbb{C}^{m\times n}$ denotes the space of $m\times n$ complex-valued matrices, $\mathbf{0}_{N}$ is the $N\times1$ zero vector, $\mathbf{I}_{N}$ represents the $N\times N$ identity matrix, $\mathbf{Diag}\{\mathbf{a}\}$ denotes a diagonal matrix having elements of $\mathbf{a}$ in order, $[\mathbf{a}]_n$ is the $n$th element in the vector $\mathbf{a}$, $\mathcal{CN}(\boldsymbol{\mu},\mathbf{\Sigma})$ is a circularly symmetric complex Gaussian random vector with the mean $\boldsymbol{\mu}$ and covariance matrix $\mathbf{\Sigma}$.

\begin{figure}[!t]
    \centering
    \includegraphics[width=3.5in]{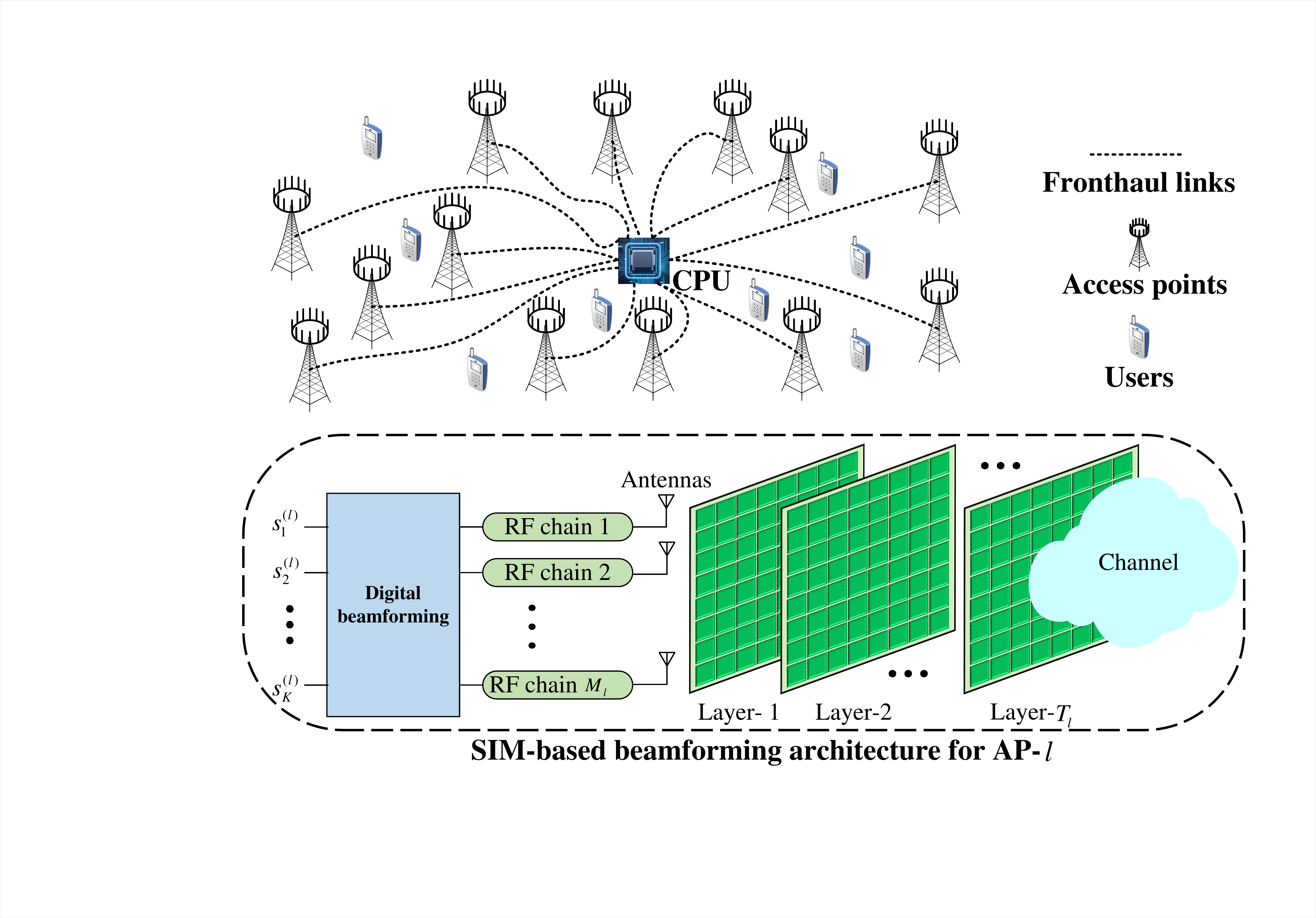}
    \caption{System model of the SIM-aided holographic MIMO in cell-free networks.}\label{Fig_system_model_HMIMO}
\end{figure}

\section{System Model}\label{System_Model}
In this section, we describe our proposed SIM-aided HMIMO architecture, and then present the channel model of the cell-free network considered.

The system model of the SIM-aided HMIMO operating in narrowband cell-free networks is shown in Fig.~\ref{Fig_system_model_HMIMO}\footnote{In this paper, we investigate the SIM-based hybrid beamforming design of narrowband cell-free networks. In wideband networks, both spatial-wideband effects and frequency-selective effects should be considered~\cite{xu2024near}. The SIM-based hybrid beamforming design of wideband networks considering the spatial-wideband effect and the frequency-selective effect is set aside for our future work.}. In contrast to the conventional cellular network, where a central AP is deployed in each cell to support the surrounding UEs, the cell-free network deploys multiple distributed APs. We focus our attention on the uplink scenario, where we consider $L$ distributed multi-antenna APs and $K$ single-antenna UEs. AP-$l$ ($l=1,2,\cdots,L$) has $M$ antennas and a stacked intelligent metasurface containing $T_l$ layers, with each layer composed of $N$ reconfigurable passive elements. Specifically, the information is sent from the $K$ UEs to $L$ APs. In each AP, the received RF signals go through the passive beamformer of the multi-layer SIM, and then they are converted to baseband signals via the RF chains. To recover the desired information, we rely on distributed operation in order to reduce the detection complexity, where each AP locally detects the data of the UEs supported by it, and then the CPU fuses the data detected at all APs via the fronthaul links to carry out the final detection of the desired information.

\subsection{SIM-Aided Holographic MIMO Architecture}
Before describing the SIM-aided holographic MIMO architecture, we briefly review the conventional fully-digital massive MIMO architecture and the hybrid digital-analog hybrid massive MIMO architecture. In the conventional fully-digital massive MIMO architecture, the number of the RF chains is the same as that of the transmit antennas. To reduce the power consumption of the RF chains, a hybrid digital-analog beamforming architecture is formed, where only a few RF chains is employed, and a phase shift array (PSA) is harnessed between the RF chains and AP antennas.

The SIM-aided HMIMO architecture of cell-free networks is shown in Fig.~\ref{Fig_system_model_HMIMO}. Firstly, the signals received at each AP go through an SIM-based beamformer. The output signals of the SIM are received by the antennas, and then converted to baseband signals via multiple RF chains. The SIM is constituted by a closed vacuum container having several layers of stacked reconfigurable metasurfaces, each of which is composed of a large number of passive densely-spaced elements~\cite{an2023stacked}, \cite{an2024stacked_earlyaccess}. By appropriately configuring the phase shifts of the elements in each layer of the metasurface with the aid of a software controllers, such as a field programmable gate array (FPGA)~\cite{an2023stacked}, \cite{an2024stacked_earlyaccess}, a substantial beamforming gain may be achieved. In practice, the SIM is enclosed in a supporting structure surrounded by wave-absorbing materials, to prevent interferences from undesired diffraction, scattering, and environmental noise~\cite{liu2022programmable}, \cite{an2024two}. The AP antennas are $M_x\times M_y$ uniform rectangular planar arrays (URPA), while the intelligent surfaces in each layer are $N_x\times N_y$ URPAs, satisfying $M=M_xM_y$ and $N=N_xN_y$. We denote the size of each reconfigurable element as $\delta_x\times\delta_y$, and the distances between the adjacent antennas are $d_x$ and $d_y$ along the $x$ and $y$ axis, respectively.

Explicitly, we contrast our SIM-aided HMIMO to various MIMO technologies in Table~\ref{Table_Comparison}.

\begin{table*}[!t]
\footnotesize
\begin{center}
\caption{Comparison of our SIM-aided HMIMO to various MIMO technologies.}
\label{Table_Comparison}
\begin{tabular}{|p{105pt}|p{130pt}|p{35pt}|p{72pt}|p{30pt}|p{36pt}|}
\hline
    & Beamforming structure & Number of RF chains & Number of PSA (or metasurface) elements & Hardware cost & Energy efficiency \\
\hline
    Fully digital MIMO & Fully digital beamformer & Large & No & High & Low \\
\hdashline
    PSA-aided hybrid MIMO & Digital beamformer + Analog beamformer relying on PSA & Moderate & Large & Moderate & Moderate \\
\hdashline
    SIM-aided HMIMO in \cite{an2023stacked}, \cite{an2024stacked_earlyaccess} & Wave-based beamformer relying on SIM & No & Very large & Low & High \\
\hdashline
    Our SIM-aided HMIMO & Digital beamformer + Wave-based beamformer relying on SIM & Moderate & Large & Low & High \\
\hline
\end{tabular}
\end{center}
\end{table*}

\subsection{Channel Model}
In this section, we describe the channel between the UEs and the AP antennas. As shown in Fig.~\ref{Fig_system_model_HMIMO}, we have to consider the channel between the SIM and the antennas at the AP as well as the channel between the UEs and the SIM at the AP.

\subsubsection{Channel links between the SIM and the antennas at APs}
We assume that the antenna array and the SIM layers are parallel to the $xoy$ plane of the Cartesian coordinate. We denote the Cartesian coordinates of the $M$ antennas at the $l$th AP as $\mathbf{p}_1^{(l,0)}=(x_1^{(l,0)},y_1^{(l,0)},z^{(l,0)})$,
$\mathbf{p}_2^{(l,0)}=(x_2^{(l,0)},y_2^{(l,0)},z^{(l,0)}),\cdots,\mathbf{p}_{M}^{(l,0)}
=(x_{M}^{(l,0)},y_{M}^{(l,0)},z^{(l,0)})$, and that of the $N$ elements in the SIM layer-$t$ as $\mathbf{p}_1^{(l,t)}=(x_1^{(l,t)},y_1^{(l,t)},z^{(l,t)})$,
$\mathbf{p}_2^{(l,t)}=(x_2^{(l,t)},y_2^{(l,t)},z^{(l,t)}),\cdots,\mathbf{p}_N^{(l,t)}
=(x_N^{(l,t)},y_N^{(l,t)},z^{(l,t)})$ with $t=1,2,\cdots,T_l$. We employ the near-field model to describe the channel response between the antennas and the SIM layers.

As shown in Fig.~\ref{Fig_system_model_HMIMO}, we denote the channel response between AP-$l$ of the SIM layer-$1$ and the antennas as $\mathbf{A}^{(l,1)}\in\mathbb{C}^{M\times N}$, with the $(m,n)$th entry $a_{m,n}^{(l,1)}$ representing the response of the $n$th element in the SIM layer-$1$ to the $m$th antenna, while $a_{m,n}^{(l,1)}$ is given by
\begin{align}\label{Channel_Model_1}
    a_{m,n}^{\left(l,1\right)}=\sqrt{\Gamma_{m,n}^{\left(l,1\right)}}
    \mathrm{e}^{-\jmath\frac{2\pi}{\lambda}
    \left\|\mathbf{p}_n^{\left(l,1\right)}-\mathbf{p}_m^{\left(l,0\right)}\right\|}.
\end{align}
In (\ref{Channel_Model_1}), $\lambda$ is the carrier wavelength, $\|\mathbf{p}_n^{(l,1)}-\mathbf{p}_m^{(l,0)}\|$ is the distance between the $n$th element in the SIM layer-$1$ and the $m$th antenna, while $\Gamma_{m,n}^{(l,1)}$ is the power radiated from the $n$th element in the SIM layer-$1$ to the $m$th antenna, which can be represented as~\cite{lu2021communicating}
\begin{align}\label{Channel_Model_2}
    \Gamma_{m,n}^{\left(l,1\right)}=\iint_{\mathcal{D}_n}
    \frac{\kappa\left(\frac{z^{\left(l,1\right)}-z^{\left(l,0\right)}}
    {\left\|\mathbf{p}_n^{\left(l,1\right)}
    -\mathbf{p}_m^{\left(l,0\right)}\right\|}\right)^{\frac{\kappa}{2}}}
    {4\pi\left\|\mathbf{p}_n^{\left(l,1\right)}-\mathbf{p}_m^{\left(l,0\right)}\right\|^2}
    \mathrm{d}x\mathrm{d}y,
\end{align}
where $\kappa$ is the directional radiation gain of the reconfigurable elements and the integration region is formulated as:
\begin{align}
    \notag\mathcal{D}_n=&\left\{\left(x,y\right)\in\mathbb{R}^2:x_n-\frac{\delta_x}{2}\leq x\leq x_n+\frac{\delta_x}{2},\right.\\
    &\left.y_n-\frac{\delta_y}{2}\leq y\leq y_n+\frac{\delta_y}{2}\right\}.
\end{align}

Next, we focus our attention on the channel model between the SIM layers. For AP-$l$ we denote the response of the channel spanning from the SIM layer-$t$ to layer-$(t-1)$ as $\mathbf{A}^{(l,t)}\in\mathbb{C}^{N\times N}$ ($t=2,3,\cdots,T_l$) with the $(n_2,n_1)$th entry representing the response from the $n_1$th element in the SIM layer-$t$ to the $n_2$th element in the SIM layer-$(t-1)$, given by
\begin{align}\label{Channel_Model_3}
    a_{n_2,n_1}^{\left(l,t\right)}=\sqrt{\Gamma_{n_2,n_1}^{\left(l,t\right)}}
    \mathrm{e}^{-\jmath\frac{2\pi}{\lambda}
    \left\|\mathbf{p}_{n_1}^{\left(l,t\right)}-\mathbf{p}_{n_2}^{\left(l,t-1\right)}\right\|}.
\end{align}
In (\ref{Channel_Model_3}), $\|\mathbf{p}_{n_1}^{(l,t)}-\mathbf{p}_{n_2}^{(l,t-1)}\|$ is the distance between the $n_1$th element in the SIM layer-$t$ and the $n_2$th element in the SIM layer-$(t-1)$, while $\Gamma_{n_2,n_1}^{(l,t)}$ is the power radiated from the $n_1$th element in the SIM layer-$t$ to the $n_2$th element in the SIM layer-$(t-1)$, which can be represented as \cite{zeng2022reconfigurable}
\begin{align}\label{Channel_Model_4_1}
    \Gamma_{n_2,n_1}^{\left(l,t\right)}=\iint_{\mathcal{D}_n}\frac{\kappa
    \left(\frac{z^{\left(l,t\right)}-z^{\left(l,t-1\right)}}
    {\left\|\mathbf{p}_{n_1}^{\left(l,t\right)}-\mathbf{p}_{n_2}^{\left(l,t-1\right)}
    \right\|}\right)^{\frac{\kappa}{2}}}
    {4\pi\left\|\mathbf{p}_{n_1}^{\left(l,t\right)}-\mathbf{p}_{n_2}^{\left(l,t-1\right)}
    \right\|^2}\mathrm{d}x\mathrm{d}y.
\end{align}

Furthermore, the phase shifts of the reconfigurable elements in each layer can be adjusted to achieve a beamforming gain. In AP-$l$, we denote the phase shift of the $n$th element in layer-$t$ as $\theta_n^{(l,t)}$. Upon considering the signal power attenuation resulting from the EM waves travelling through a SIM, we denote the radiation coefficient of the $n$th element in layer-$t$ as $\upsilon_n^{(l,t)}$ satisfying $\upsilon_n^{(l,t)}\in[0,1]$. Thus, by defining $\mathbf{\Upsilon}^{(l,t)}=\mathbf{Diag}\{\upsilon_1^{(l,t)},
\upsilon_2^{(l,t)},\cdots,\upsilon_N^{(l,t)}\}$ and $\mathbf{\Theta}^{(l,t)}=\mathbf{Diag}\{\mathrm{e}^{\jmath\theta_1^{(l,t)}},
\mathrm{e}^{\jmath\theta_2^{(l,t)}},\cdots,\mathrm{e}^{\jmath\theta_N^{(l,t)}}\}$, the response matrix of the SIM's layer-$t$ can be represented as
\begin{align}
    \notag\mathbf{\Xi}^{\left(l,t\right)}=
    &\sqrt{\mathbf{\Upsilon}^{\left(l,t\right)}}\mathbf{\Theta}^{\left(l,t\right)}\\
    \notag=&\mathbf{Diag}\left\{\sqrt{\upsilon_1^{\left(l,t\right)}}
    \mathrm{e}^{\jmath\theta_1^{\left(l,t\right)}},
    \sqrt{\upsilon_2^{\left(l,t\right)}}\mathrm{e}^{\jmath\theta_2^{\left(l,t\right)}},
    \cdots,\right.\\
    &\left.\sqrt{\upsilon_N^{\left(l,t\right)}}\mathrm{e}^{\jmath\theta_N^{\left(l,t\right)}}\right\}.
\end{align}

Therefore, the equivalent channel response of the SIM-based beamformer at the $l$th AP, denoted as $\mathbf{G}^{(l)}\in\mathbb{C}^{M\times N}$, is\footnote{In this paper, we assume that each SIM layer can be configured independently without mutual coupling. In reality, having mutual coupling among the metamaterial elements is inevitable due to their dense packaging. The holographic beamforming design considering the effect of mutual coupling is set aside for our future work.}
\begin{align}\label{Channel_Model_4_2}
    \notag\mathbf{G}^{(l)}=&\mathbf{A}^{\left(l,1\right)}\mathbf{\Xi}^{\left(l,1\right)}
    \mathbf{A}^{\left(l,2\right)}\mathbf{\Xi}^{\left(l,2\right)}
    \cdots\mathbf{A}^{\left(l,T_l\right)}\mathbf{\Xi}^{\left(l,T_l\right)}\\
    \notag=&\mathbf{A}^{\left(l,1\right)}
    \sqrt{\mathbf{\Upsilon}^{\left(l,1\right)}}\mathbf{\Theta}^{\left(l,1\right)}
    \mathbf{A}^{\left(l,2\right)}
    \sqrt{\mathbf{\Upsilon}^{\left(l,2\right)}}\mathbf{\Theta}^{\left(l,2\right)}
    \cdots\mathbf{A}^{\left(l,T_l\right)}\\
    &\sqrt{\mathbf{\Upsilon}^{\left(l,T_l\right)}}\mathbf{\Theta}^{\left(l,T_l\right)}.
\end{align}

\subsubsection{Channel links between UEs and APs}
We denote the large-scale fading and the small-scale fading between UE-$k$ and the SIM at AP-$l$ as $\varrho_k^{(l)}$ and $\mathbf{h}_k^{(l)}\in\mathbb{C}^{N\times1}$, respectively. We assume the knowledge of $\mathbf{h}_1^{(l)},\mathbf{h}_2^{(l)},\cdots,\mathbf{h}_K^{(l)}$ can be attained at the AP-$l$. In practical metasurface-based holographic MIMO systems, this has to be acquired by CSI acquisition methods, such as the subspace-based channel estimator of~\cite{demir2022channel} or the sparse channel estimator of~\cite{cui2022channel}. We adopt the millimeter-wave (mmWave) channel model to characterize the propagation environment between the APs and each user\footnote{Note that here we consider a mmWave channel model as an example for describing the signal propagation between the APs and the UEs. But indeed, our proposed SIM-based beamforming architecture is also applicable to other channel models.}. Specifically, the mmWave uplink channel of UE-$k$ is assumed to be the superposition of all propagation paths that are scattered in $\zeta_{c}$ clusters and each cluster contributes $\zeta_{p}$ paths, expressed as~\cite{hemadeh2017millimeter}
\begin{align}\label{Channel_Model_5}
    \mathbf{h}_k^{(l)}=\sqrt{\frac{1}{\zeta_c\zeta_p}}
    \sum_{c=1}^{\zeta_c}\sum_{p=1}^{\zeta_p}\alpha_{c,p}^{\left(l,k\right)}
    \mathbf{f}(\psi_{c,p}^{\left(l,k\right)},\varphi_{c,p}^{\left(l,k\right)}),
\end{align}
where $\alpha_{c,p}^{(l,k)}$ is the complex gain of the $p$th path in the $c$th cluster following $\alpha_{c,p}^{(l,k)}\sim\mathcal{CN}(0,1)$, and $\mathbf{f}(\psi_{c,p}^{(l,k)},\varphi_{c,p}^{(l,k)})$ is
\begin{align}\label{Channel_Model_6}
    \notag&\mathbf{f}\left(\psi_{c,p}^{\left(l,k\right)},\varphi_{c,p}^{\left(l,k\right)}\right)
    =\left[1,\cdots,\right.\\
    \notag&\left.\mathrm{e}^{\jmath\frac{2\pi}{\lambda}
    \left(\delta_xn_x\sin\psi_{c,p}^{\left(l,k\right)}\cos\varphi_{c,p}^{\left(l,k\right)}
    +\delta_yn_y\sin\psi_{c,p}^{\left(l,k\right)}\sin\varphi_{c,p}^{\left(l,k\right)}\right)},
    \cdots\right.\\
    &\left.\mathrm{e}^{\jmath\frac{2\pi}{\lambda}
    \left(\delta_x\left(N_x-1\right)\sin\psi_{c,p}^{\left(l,k\right)}
    \cos\varphi_{c,p}^{\left(l,k\right)}
    +\delta_y\left(N_y-1\right)\sin\psi_{c,p}^{\left(l,k\right)}
    \sin\varphi_{c,p}^{\left(l,k\right)}\right)}\right]^\mathrm{H},
\end{align}
where $\psi_{c,p}^{(l,k)}$ and $\varphi_{c,p}^{(l,k)}$ are the elevation and azimuth angles of departure (AoD) from UE-$k$ to the SIM in AP-$l$ in the $p$th path of the $c$th cluster, respectively. Within the $c$th cluster, the random variables $\psi_{c,p}^{(l,k)}$ and $\varphi_{c,p}^{(l,k)}$ have uniformly distributed mean values of $\mu_{\psi_{c}^k}$ and $\mu_{\varphi_{c}^k}$, respectively, and have angular spreads, i.e. standard deviations, of $\sigma_{\psi_{c}^{k}}$ and $\sigma_{\varphi_{c}^{k}}$, respectively.

The signal received by the antennas at AP-$l$ is given by
\begin{align}\label{Channel_Model_7}
    \notag\mathbf{y}^{(l)}=&\sum_{k=1}^{K}\left(\sqrt{\rho_k\varepsilon_{u_k}
    \varepsilon_{\mathbf{v}^{(l)}}}\mathbf{q}_k^{(l)}s_k+\sqrt{\rho_k
    \left(1-\varepsilon_{u_k}\right)\varepsilon_{\mathbf{v}^{(l)}}}\mathbf{q}_k^{(l)}u_k\right.\\
    &\left.+\sqrt{\rho_k(1-\varepsilon_{\mathbf{v}^{(l)}})}\mathbf{q}_k^{(l)}
    \odot\mathbf{v}_k^{(l)}\right)+\mathbf{w}^{(l)},
\end{align}
where $\mathbf{q}_k^{(l)}=\sqrt{\varrho_k^{(l)}}\mathbf{G}^{(l)}\mathbf{h}_k^{(l)}$ is the equivalent channel spanning from the UE-$k$ to the antennas at AP-$l$, $s_k$ is the desired information of UE-$k$, $\rho_k$ denotes the transmit power of UE-$k$, and $\mathbf{w}^{(l)}\sim\mathcal{CN}(\mathbf{0}_{M},\sigma_{w^{(l)}}^2\mathbf{I}_{M})$ is the additive noise at AP-$l$. Furthermore, $u_k\sim\mathcal{CN}(0,1)$ represents the contamination of the information symbol $s_k$ due to HWIs at UE-$k$, resulting from the power amplifier non-linearities, amplitude/phase imbalance in the In-phase/Quadrature mixers, phase noise in the local oscillator, sampling jitter and finite-resolution quantization in the analog-to-digital converters. Furthermore, $\mathbf{v}_k^{(l)}\sim\mathcal{CN}(\mathbf{0}_{M},\mathbf{I}_{M})$ is the distortion of the information symbol $s_k$ due to HWIs of the RF chains at AP-$l$. Finally, $\varepsilon_{u_k}$ and $\varepsilon_{\mathbf{v}^{(l)}}$ represents the hardware quality factors of UE-$k$ and AP-$l$ satisfying $0\leq\varepsilon_{u_k}\leq1$ and $0\leq\varepsilon_{\mathbf{v}^{(l)}}\leq1$, respectively~\cite{bjornson2017massive,
li2023achievable}. Explicitly, a hardware quality factor of 1 indicates that the hardware is ideal, while 0 means that the hardware is completely inadequate.

\section{Beamforming Design}\label{Beamforming_Design}
As shown in the SIM-based cell-free network of Fig.~\ref{Fig_system_model_HMIMO}, we rely on the distributed processing philosophy for reducing the required overhead of CSI sharing between APs. Specifically, for the information transmitted from the UE-$k$, each AP carries out a local detection based on its received signal, where the SIM coefficient matrices and the RC vectors of each AP are optimized based on the corresponding local CSI. Then, the CPU gathers the locally detected signals from all APs to generate a final estimate of the UE-$k$ information. In the following, we first present the beamformer design in terms of the SIM coefficient matrices and combining vectors at each AP, followed by the optimization of the weight vectors at the CPU for generating the final detection of the UE information.

\subsection{SIM Coefficient Matrices and RC Vectors Design at APs}
To recover the information transmitted from UE-$k$, we denote the corresponding RC vector at AP-$l$ as $\mathbf{b}_k^{(l)}$ satisfying $\|\mathbf{b}_k^{(l)}\|^2=1$. Thus, the locally recovered information of $s_k$ at AP-$l$, denoted as $\hat{s}_k^{(l)}$, is given by
\begin{align}\label{Beamforming_Design_8}
    \notag\hat{s}_k^{(l)}=&\mathbf{b}_k^{(l)\mathrm{H}}\mathbf{y}^{(l)}\\
    \notag=&\sum_{k=1}^{K}
    \left(\sqrt{\rho_k\varepsilon_{u_k}\varepsilon_{\mathbf{v}^{(l)}}}
    \mathbf{b}_k^{(l)\mathrm{H}}\mathbf{q}_k^{(l)}s_k\right.\\
    \notag&\left.+\sqrt{\rho_k\left(1-\varepsilon_{u_k}\right)\varepsilon_{\mathbf{v}^{(l)}}}
    \mathbf{b}_k^{(l)\mathrm{H}}\mathbf{q}_k^{(l)}u_k\right.\\
    &\left.+\sqrt{\rho_k\left(1-\varepsilon_{\mathbf{v}^{(l)}}\right)}
    \mathbf{b}_k^{(l)\mathrm{H}}\left(\mathbf{q}_k^{(l)}\odot\mathbf{v}_k^{(l)}\right)\right)
    +\mathbf{b}_k^{(l)\mathrm{H}}\mathbf{w}^{(l)},
\end{align}
and the corresponding signal-to-interference-plus-noise ratio (SINR), denoted as $\gamma_k^{(l)}$, can be formulated as
\begin{align}\label{Beamforming_Design_9}
    \gamma_k^{(l)}=\frac{\rho_k\varepsilon_{u_k}\varepsilon_{\mathbf{v}^{(l)}}
    \left|\mathbf{b}_k^{(l)\mathrm{H}}\mathbf{q}_k^{(l)}\right|^2}
    {\varsigma_{k,k}+\mathop{\sum\limits_{k'=1}^{K}}\limits_{k'\neq k}
    \left(\rho_{k'}\varepsilon_{u_{k'}}\varepsilon_{\mathbf{v}^{(l)}}
    \left|\mathbf{b}_k^{(l)\mathrm{H}}\mathbf{q}_{k'}^{(l)}\right|^2+\varsigma_{k,k'}\right)
    +\sigma_{w^{(l)}}^2}.
\end{align}
In (\ref{Beamforming_Design_9}) $\varsigma_{k,{k'}}$ represents the interference resulting from the distortion imposed on UE-$k'$ signal for the recovery of $s_k$ due to the HWIs, given by
\begin{align}
    \notag\varsigma_{k,{k'}}=&\rho_{k'}\left(1-\varepsilon_{u_{k'}}\right)\varepsilon_{\mathbf{v}^{(l)}}
    \left|\mathbf{b}_k^{(l)\mathrm{H}}\mathbf{q}_{k'}^{(l)}\right|^2
    +\rho_{k'}\left(1-\varepsilon_{\mathbf{v}^{(l)}}\right)\cdot\\
    &\mathbf{b}_k^{(l)\mathrm{H}}\left(\left(\mathbf{q}_{k'}^{(l)}
    \mathbf{q}_{k'}^{(l)\mathrm{H}}\right)\odot\mathbf{I}_{M}\right)\mathbf{b}_k^{(l)}.
\end{align}

For a specific UE-$k$, we aim for jointly optimizing the active beamformers $\mathbf{b}_1^{(l)},\mathbf{b}_2^{(l)},\cdots,\mathbf{b}_K^{(l)}$ and the passive SIM-based beamformers $\mathbf{\Theta}^{(l,1)},\mathbf{\Theta}^{(l,2)},\cdots,
\mathbf{\Theta}^{(l,T_l)}$ to maximize the SINR of $\gamma_k^{(l)}$ with $l=1,2,\cdots,L$. The corresponding optimization problem can be formulated as
\begin{align}
    \text{(P1)}&\max_{\mathbf{b}_1^{(l)},\mathbf{b}_2^{(l)},\cdots,\mathbf{b}_K^{(l)},
    \mathbf{\Theta}^{\left(l,1\right)},\mathbf{\Theta}^{\left(l,2\right)},\cdots,
    \mathbf{\Theta}^{\left(l,T_l\right)}}\gamma_k^{(l)},\  l=1,2,\cdots,L\\
    \text{s.t.}&\quad \mathbf{\Theta}^{\left(l,t\right)}
    \mathbf{\Theta}^{\left(l,t\right)\mathrm{H}}=\mathbf{I}_N,\quad t=1,2,\cdots,T_l,\\
    &\quad \left\|\mathbf{b}_{k'}^{(l)}\right\|^2=1,\quad k'=1,2,\cdots,K.
\end{align}
Since the $L$ APs jointly support all $K$ UEs, the beamforming of all APs focused on a specific UE-$k$ to maximize the SINR $\gamma_k^{(l)}$ ($l=1,2,\cdots,L$) is carried out at the cost of disregarding the other $K-1$ UEs. The main advantage of the cell-free architecture is that of reducing the path-loss from each AP to its nearest UE. Thus, at each AP we aim for maximizing the channel gain between the AP and its nearest user within the SIM-based holographic beamformer, while the digital beamformer is optimized for all $K$ UEs to get their local information estimates. Specifically, we denote the AP set associated with the SIM-based beamforming focused on UE-$k$ as $\mathcal{L}_k=\{l:D_k^{(l)}\leq D_k^{(l')},l'=1,2,\cdots,L\}$ with $D_k^{(l)}$ being the distance from UE-$k$ to AP-$l$. Therefore, problem (P1) can be formulated as
\begin{align}
    \text{(P2)}&\max_{\mathbf{b}_1^{(l)},\mathbf{b}_2^{(l)},\cdots,\mathbf{b}_K^{(l)},
    \mathbf{\Theta}^{\left(l,1\right)},\mathbf{\Theta}^{\left(l,2\right)},\cdots,
    \mathbf{\Theta}^{\left(l,T_l\right)}}\gamma_k^{(l)},\ l\in\mathcal{L}_k\\
    \text{s.t.}&\quad\mathbf{\Theta}^{\left(l,t\right)}\mathbf{\Theta}^{\left(l,t\right)\mathrm{H}}
    =\mathbf{I}_N,\quad t=1,2,\cdots,T_l,\\
    &\quad \left\|\mathbf{b}_{k'}^{(l)}\right\|^2=1,\quad k'=1,2,\cdots,K.
\end{align}
Since (P2) is a non-convex problem, we can decouple it into a pair of sub-problems and optimize them iteratively as follows.

\subsubsection{Design of RC vectors at APs}
Once the SIM-based beamformer $\mathbf{\Theta}^{(l,1)},\mathbf{\Theta}^{(l,2)},\cdots,
\mathbf{\Theta}^{(l,T_l)}$ is given, we can estimate the equivalent channels impinging from all $K$ UEs to AP-$l$, i.e $\mathbf{q}_1^{(l)},\mathbf{q}_2^{(l)},\cdots,\mathbf{q}_K^{(l)}$. Then, the optimal active beamformer $\mathbf{b}_{k'}^{(l)}$ designed for the recovery of the information $s_{k'}$ at the AP-$l$ can be optimized by using the MRC criterion as
\begin{align}
    \mathbf{b}_{k'}^{(l)}=\frac{\mathbf{q}_{k'}^{(l)}}{\left\|\mathbf{q}_{k'}^{(l)}\right\|},
    \quad k'=1,2,\cdots,K.
\end{align}

\subsubsection{Design of SIM coefficient matrices at APs}
When the active beamformer $\mathbf{b}_k^{(l)}$ is given, the problem (P2) aims for optimizing the channel gain $\|\mathbf{G}^{(l)}\mathbf{h}_k^{(l)}\|^2$, and it is given by
\begin{align}
    \text{(P3)}\quad &\max_{\mathbf{\Theta}_1^{(l)},\mathbf{\Theta}_2^{(l)},\cdots,
    \mathbf{\Theta}_{T_l}^{(l)}}\left\|\mathbf{G}^{(l)}\mathbf{h}_k^{(l)}\right\|^2,\  l\in\mathcal{L}_k\\
    \text{s.t.}&\quad \mathbf{\Theta}_{t}^{(l)}\mathbf{\Theta}_{t}^{(l)\mathrm{H}}=\mathbf{I}_N,
    \quad t=1,2,\cdots,T_l.
\end{align}
Since the sub-problem (P3) is still a non-convex one, we propose a layer-by-layer iterative optimization algorithm. Firstly, we optimize $\mathbf{\Theta}^{(l,1)}$ by fixing all the other $T_l-1$ layers of the SIM. Therefore, the channel gain can be represented as $\overline{\mathbf{b}}_k^{(l,1)\mathrm{H}}\mathbf{\Theta}^{(l,1)}\overline{\mathbf{h}}_k^{(l,1)}$
in conjunction with
\begin{align}
    \overline{\mathbf{b}}_k^{\left(l,1\right)\mathrm{H}}=\mathbf{b}_k^{(l)\mathrm{H}}
    \mathbf{A}^{\left(l,1\right)}\sqrt{\mathbf{\Upsilon}^{\left(l,1\right)}}
\end{align}
and
\begin{align}
    \overline{\mathbf{h}}_k^{\left(l,1\right)}=\mathbf{A}^{\left(l,2\right)}
    \sqrt{\mathbf{\Upsilon}^{\left(l,2\right)}}\mathbf{\Theta}^{\left(l,2\right)}\cdots
    \mathbf{A}^{\left(l,T_l\right)}\sqrt{\mathbf{\Upsilon}^{\left(l,T_l\right)}}
    \mathbf{\Theta}^{\left(l,T_l\right)}\mathbf{h}_k^{(l)},
\end{align}
and the passive beamformer $\mathbf{\Theta}^{(l,1)}$ can be optimized as
\begin{align}
    \mathbf{\Theta}^{\left(l,1\right)}=\mathbf{Diag}
    \left\{\mathrm{e}^{\jmath\left(\angle\overline{\mathbf{b}}_k^{\left(l,1\right)}
    -\angle\overline{\mathbf{h}}_k^{\left(l,1\right)}\right)}\right\}.
\end{align}
Afterwards, we optimize $\mathbf{\Theta}^{(l,2)}$ by fixing all the other $T_l-1$ layers of the SIM with the channel gain represented as $\overline{\mathbf{b}}_k^{(l,2)\mathrm{H}}\mathbf{\Theta}^{(l,2)}
\overline{\mathbf{h}}_k^{(l,2)}$ along with
\begin{align}
    \overline{\mathbf{b}}_k^{\left(l,2\right)\mathrm{H}}
    =\mathbf{b}_k^{(l)\mathrm{H}}\mathbf{A}^{\left(l,1\right)}
    \sqrt{\mathbf{\Upsilon}^{\left(l,1\right)}}\mathbf{\Theta}^{\left(l,1\right)}
    \mathbf{A}^{\left(l,2\right)}\sqrt{\mathbf{\Upsilon}^{\left(l,2\right)}}
\end{align}
and
\begin{align}
    \overline{\mathbf{h}}_k^{\left(l,2\right)}=\mathbf{A}^{\left(l,3\right)}
    \sqrt{\mathbf{\Upsilon}^{\left(l,3\right)}}\mathbf{\Theta}^{\left(l,3\right)}\cdots
    \mathbf{A}^{\left(l,T_l\right)}\sqrt{\mathbf{\Upsilon}^{\left(l,T_l\right)}}
    \mathbf{\Theta}^{\left(l,T_l\right)}\mathbf{h}_k^{(l)},
\end{align}
and the passive beamformer $\mathbf{\Theta}^{(l,2)}$ can be optimized as
\begin{align}
    \mathbf{\Theta}^{\left(l,2\right)}=\mathbf{Diag}
    \left\{\mathrm{e}^{\jmath\left(\angle\overline{\mathbf{b}}_k^{\left(l,2\right)}
    -\angle\overline{\mathbf{h}}_k^{(l,2)}\right)}\right\}.
\end{align}
Then, $\mathbf{\Theta}^{(l,3)},\mathbf{\Theta}^{(l,4)},\cdots,\mathbf{\Theta}^{(l,T_l)}$ can be optimized in turn.

The details of the layer-by-layer iterative optimization of the hybrid beamformer is presented in Algorithm~\ref{algorithm_1}.

\begin{algorithm}[!t]
\caption{Layer-by-layer iterative optimization algorithm for hybrid beamforming at AP-$l$ ($l\in\mathcal{L}_k$).}
\label{algorithm_1}
\begin{algorithmic}[1]
\REQUIRE
    The power attenuation matrices of SIM elements $\mathbf{\Upsilon}^{(l,1)},\mathbf{\Upsilon}^{(l,2)},\cdots,
    \mathbf{\Upsilon}^{(l,T_l)}$, and the channel links $\mathbf{A}^{(l,1)},\mathbf{A}^{(l,2)},\cdots,
    \mathbf{A}^{(l,T_l)}$ and $\mathbf{h}_k^{(l)}$.
    \STATE
        Set the random initial passive SIM-based beamforming matrices $\mathbf{\Theta}^{(l,1)}$, $\mathbf{\Theta}^{(l,2)}$, $\cdots$, $\mathbf{\Theta}^{(l,T_l)}$, satisfying $\mathbf{\Theta}^{(l,1)}\mathbf{\Theta}^{(l,1)\mathrm{H}}
        =\mathbf{\Theta}^{(l,2)}\mathbf{\Theta}^{(l,2)\mathrm{H}}
        =\mathbf{\Theta}^{(l,T_l)}\mathbf{\Theta}^{(l,T_l)\mathrm{H}}=\mathbf{I}_N$.
    \REPEAT
        \STATE
            The equivalent channel from the UE-$k$ to AP-$l$ antennas is $\mathbf{q}_k^{(l)}=\sqrt{\varrho_k^{(l)}}\mathbf{A}^{(l,1)}\sqrt{\mathbf{\Upsilon}^{(l,1)}}
            \mathbf{\Theta}^{(l,1)}\mathbf{A}^{(l,2)}\sqrt{\mathbf{\Upsilon}^{(l,2)}}\newline
            \mathbf{\Theta}^{(l,2)}\cdots\mathbf{A}^{(l,T_l)}
            \sqrt{\mathbf{\Upsilon}^{(l,T_l)}}\mathbf{\Theta}^{(l,T_l)}\mathbf{h}_k^{(l)}$.
        \STATE
            Design the active beamforming vector by the MRC method as $\mathbf{b}_{k'}^{(l)}=\frac{\mathbf{q}_{k'}^{(l)}}
            {\|\mathbf{q}_{k'}^{(l)}\|}$ for $k'=1,2,\cdots,K$.
        \FOR{$t=1$ to $T_l$}
            \STATE $\overline{\mathbf{b}}_k^{(l,t)}
            =\mathbf{b}_k^{(l)\mathrm{H}}\mathbf{A}^{(l,1)}\sqrt{\mathbf{\Upsilon}^{(l,1)}}
            \mathbf{\Theta}^{(l,1)}\cdots\mathbf{A}^{(l,t)}\sqrt{\mathbf{\Upsilon}^{(l,t)}}$.
            \STATE
            $\overline{\mathbf{h}}_k^{(l,t)}=\mathbf{A}^{(l,t+1)}\sqrt{\mathbf{\Upsilon}^{(l,t+1)}}
            \mathbf{\Theta}^{(l,t+1)}\cdots\mathbf{A}^{(l,T_l)}\newline
            \sqrt{\mathbf{\Upsilon}^{(l,T_l)}}\mathbf{\Theta}^{(l,T_l)}\mathbf{h}_k^{(l)}$.
        \STATE
            The optimal SIM-based passive beamforming matrix for layer-$t$ is $\mathbf{\Theta}^{(l,t)}=\mathbf{Diag}\{{\mathrm{e}^{\jmath(\angle
            \overline{\mathbf{b}}_k^{(l,t)}-\angle\overline{\mathbf{h}}_k^{(l,t)})}}\}$.
        \ENDFOR
    \UNTIL{reaching the iteration times.}
\ENSURE
    The optimized active beamforming vectors $\mathbf{b}_k^{(l)}$ and the optimized passive SIM-based beamforming matrices $\mathbf{\Theta}^{(l,1)},\mathbf{\Theta}^{(l,2)},\cdots,
    \mathbf{\Theta}^{(l,T_l)}$.
\end{algorithmic}
\end{algorithm}

\subsection{Weight Vector Design at the CPU}
The CPU computes its estimate as a linear combination of the local estimates as~\cite{demir2021foundations}
\begin{align}\label{Beamforming_CPU}
    \hat{s}_k=\sum_{l=1}^L\eta_k^{(l)\dag}\hat{s}_k^{(l)},
\end{align}
where $\boldsymbol{\eta}_k=[\eta_k^{(l)},\eta_k^{(2)},\cdots,\eta_k^{(L)}]^\mathrm{T}
\in\mathbb{C}^{L\times1}$ is the weight vector that the CPU assigns to the local signal estimate of the signal arriving from UE-$k$ satisfying $\|\boldsymbol{\eta}_k\|^2=1$. Therefore, according to (\ref{Beamforming_Design_8}), $\hat{s}_k$ can be written as in (\ref{Beamforming_Design_10}).
\begin{figure*}
\begin{align}\label{Beamforming_Design_10}
    \notag\hat{s}_k
    =&\sum_{l=1}^L\eta_k^{(l)\dag}\mathbf{b}_k^{(l)\mathrm{H}}\mathbf{y}^{(l)}\\
    \notag=&\underbrace{\sum_{l=1}^L\eta_k^{(l)\dag}\sqrt{\rho_k\varepsilon_{u_k}
    \varepsilon_{\mathbf{v}^{(l)}}}
    \left\|\mathbf{q}_k^{(l)}\right\|s_k}_{\text{Desired signal for $s_k$}}
    +\underbrace{\sum_{l=1}^L\eta_k^{(l)\dag}\sqrt{\rho_k\left(1-\varepsilon_{u_k}\right)
    \varepsilon_{\mathbf{v}^{(l)}}}
    \left\|\mathbf{q}_k^{(l)}\right\|u_k}_{\text{Signal distortion resulting from the HWI at UE-$k$}}
    +\underbrace{\sum_{l=1}^L\eta_k^{(l)\dag}\sqrt{\rho_k\left(1-\varepsilon_{\mathbf{v}^{(l)}}
    \right)}
    \frac{\mathbf{q}_k^{(l)\mathrm{H}}
    \left(\mathbf{q}_k^{(l)}\odot\mathbf{v}_k^{(l)}\right)}
    {\left\|\mathbf{q}_k^{(l)}\right\|}}_{\text{Signal distortion resulting from the HWI at APs}}\\
    \notag&+\underbrace{\mathop{\sum_{k'=1}^{K}}_{k'\neq k}\sum_{l=1}^L\eta_k^{(l)\dag}
    \Bigg(\sqrt{\rho_{k'}\varepsilon_{u_{k'}}\varepsilon_{\mathbf{v}^{(l)}}}
    \frac{\mathbf{q}_k^{(l)\mathrm{H}}\mathbf{q}_{k'}^{(l)}s_{k'}}
    {\left\|\mathbf{q}_k^{(l)}\right\|}
    +\sqrt{\rho_{k'}(1-\varepsilon_{u_{k'}})\varepsilon_{\mathbf{v}^{(l)}}}
    \frac{\mathbf{q}_k^{(l)\mathrm{H}}\mathbf{q}_{k'}^{(l)}u_{k'}}
    {\left\|\mathbf{q}_k^{(l)}\right\|}+\sqrt{\rho_{k'}
    \left(1-\varepsilon_{\mathbf{v}^{(l)}}\right)}\frac{\mathbf{q}_k^{(l)\mathrm{H}}
    \left(\mathbf{q}_{k'}^{(l)}\odot\mathbf{v}_{k'}^{(l)}\right)}
    {\left\|\mathbf{q}_k^{(l)}\right\|}\Bigg)}_{\text{Inter-user interference}}\\
    &+\underbrace{\sum_{l=1}^L\eta_k^{(l)\dag}
    \frac{\mathbf{q}_k^{(l)\mathrm{H}}\mathbf{w}^{(l)}}{\left\|\mathbf{q}_k^{(l)}\right\|}}
    _{\text{Additive noise}}
\end{align}
\hrulefill
\end{figure*}
The SINR of $\hat{s}_k$, denoted as $\gamma_k$, can be derived as
\begin{align}\label{Beamforming_Design_11}
    \gamma_k=\frac{\rho_k\left|\boldsymbol{\eta}_k^\mathrm{H}\mathbf{z}_{k,k}\right|^2}
    {\boldsymbol{\eta}_k^\mathrm{H}\mathbf{R}_k\boldsymbol{\eta}_k},
\end{align}
where $\mathbf{R}_k$ is given by
\begin{align}\label{Beamforming_Design_12}
    \notag\mathbf{R}_k=&\rho_k\left(\mathbf{z}_{u_{k,k}}\mathbf{z}_{u_{k,k}}^\mathrm{H}
    +\left(\mathbf{z}_{\mathbf{v}_{k,k}}\mathbf{z}_{\mathbf{v}_{k,k}}^\mathrm{H}\right)
    \odot\mathbf{I}_L\right)+\mathop{\sum_{k'=1}^{K}}_{k'\neq k}\rho_{k'}\left(\mathbf{z}_{k,k'}\right.\\
    &\left.\mathbf{z}_{k,k'}^\mathrm{H}
    +\mathbf{z}_{u_{k,k'}}\mathbf{z}_{u_{k,k'}}^\mathrm{H}
    +\left(\mathbf{z}_{\mathbf{v}_{k,k'}}\mathbf{z}_{\mathbf{v}_{k,k'}}^\mathrm{H}\right)
    \odot\mathbf{I}_L\right)+\mathbf{W}
\end{align}
in conjunction with
\begin{align}\label{Beamforming_Design_12_1}
    \mathbf{z}_{k,i}=
    \left[\begin{array}{c}
        \frac{\sqrt{\varepsilon_{u_k}\varepsilon_{\mathbf{v}^{(1)}}}}
        {\left\|\mathbf{q}_k^{(1)}\right\|}
        \mathbf{q}_k^{(1)\mathrm{H}}\mathbf{q}_i^{(1)} \\
        \vdots \\
        \frac{\sqrt{\varepsilon_{u_k}\varepsilon_{\mathbf{v}^{(L)}}}}
        {\left\|\mathbf{q}_k^{(L)}\right\|}
        \mathbf{q}_k^{(L)\mathrm{H}}\mathbf{q}_i^{(L)}
    \end{array}\right],
\end{align}
\begin{align}\label{Beamforming_Design_12_2}
    \mathbf{z}_{u_{k,i}}=
    \left[\begin{array}{c}
        \frac{\sqrt{\left(1-\varepsilon_{u_k}\right)\varepsilon_{\mathbf{v}^{(1)}}}}
        {\left\|\mathbf{q}_k^{(1)}\right\|}
        \mathbf{q}_k^{(1)\mathrm{H}}\mathbf{q}_i^{(1)} \\
        \vdots \\
        \frac{\sqrt{\left(1-\varepsilon_{u_k}\right)\varepsilon_{\mathbf{v}^{(L)}}}}
        {\left\|\mathbf{q}_k^{(L)}\right\|}
        \mathbf{q}_k^{(L)\mathrm{H}}\mathbf{q}_i^{(L)}
    \end{array}\right],
\end{align}
\begin{align}\label{Beamforming_Design_12_3}
    \mathbf{z}_{\mathbf{v}_{k,i}}=
    \left[\begin{array}{c}
        \frac{\sqrt{1-\varepsilon_{\mathbf{v}^{(1)}}}}{\left\|\mathbf{q}_k^{(L)}\right\|}
        \left\|\mathbf{q}_k^{(1)\dag}\odot\mathbf{q}_i^{(1)}\right\| \\
        \vdots \\
        \frac{\sqrt{1-\varepsilon_{\mathbf{v}^{(L)}}}}{\left\|\mathbf{q}_k^{(L)}\right\|}
        \left\|\mathbf{q}_k^{(L)\dag}\odot\mathbf{q}_i^{(L)}\right\|
    \end{array}\right],
\end{align}
and
\begin{align}\label{Beamforming_Design_12_4}
    \mathbf{W}=\mathbf{Diag}\{\sigma_{w^{(1)}}^2,\sigma_{w^{(2)}}^2,\cdots,
    \sigma_{w^{(L)}}^2\}.
\end{align}
Based on the generalized Rayleigh quotient, the maximum of $\gamma_k$ in (\ref{Beamforming_Design_11}) can be attained as follows:
\begin{align}\label{Beamforming_Design_13}
    \notag\gamma_k=&\rho_k\mathbf{z}_{k,k}^\mathrm{H}\mathbf{R}_k^{-1}\mathbf{z}_{k,k}\\
    \notag=&\rho_k\mathbf{z}_{k,k}^\mathrm{H}\Big(\rho_k\left(\mathbf{z}_{u_{k,k}}
    \mathbf{z}_{u_{k,k}}^\mathrm{H}+\left(\mathbf{z}_{\mathbf{v}_{k,k}}
    \mathbf{z}_{\mathbf{v}_{k,k}}^\mathrm{H}\right)\odot\mathbf{I}_L\right)\\
    \notag&+\mathop{\sum_{k'=1}^{K}}_{k'\neq k}
    \rho_{k'}\left(\mathbf{z}_{k,k'}\mathbf{z}_{k,k'}^\mathrm{H}
    +\mathbf{z}_{u_{k,k'}}\mathbf{z}_{u_{k,k'}}^\mathrm{H}\right.\\
    &\left.+\left(\mathbf{z}_{\mathbf{v}_{k,k'}}\mathbf{z}_{\mathbf{v}_{k,k'}}^\mathrm{H}\right)
    \odot\mathbf{I}_L\right)+\mathbf{W}\Big)^{-1}\mathbf{z}_{k,k},
\end{align}
by setting
\begin{align}\label{Beamforming_Design_14}
    \boldsymbol{\eta}_k=\mathbf{R}_k^{-1}\mathbf{z}_{k,k}.
\end{align}

\subsection{Computational Complexity}
\subsubsection{Computational complexity of the hybrid beamformer at each AP}
The computational complexity of the proposed layer-by-layer iterative hybrid beamformer at AP-$l$ ($l\in\mathcal{L}_k$) in Algorithm~\ref{algorithm_1} depends both on the number of iterations in the alternating maximization, which is denoted as $\tau$, and on the computational complexity required in each iteration. As for each iteration, the sub-problem of optimizing the combining vectors $\mathbf{b}_1^{(l)},\mathbf{b}_2^{(l)},\cdots,\mathbf{b}_K^{(l)}$ is solved in line 3 and 4 of Algorithm~\ref{algorithm_1}, while the sub-problem of optimizing the coefficients of all layers in the SIM $\mathbf{\Theta}^{(l,1)},\mathbf{\Theta}^{(l,2)},\cdots,
\mathbf{\Theta}^{(l,T_l)}$ is evaluated in line 5 to 9 of Algorithm~\ref{algorithm_1}. The complexity of additions is neglected, since its operation is easily implemented in hardware. Hence, we quantify the computational complexity by counting the number of floating-point multiplication operations that are required. Specifically, the calculation of $\mathbf{q}_k^{(l)}$ in line 3 of Algorithm~\ref{algorithm_1} requires
\begin{align}\label{Complexity_1}
    c_1'=\left(T_l-1\right)\left(N^2+2N\right)+MN+2N+M
\end{align}
floating-point multiplication operations by exploiting the property that $\mathbf{\Upsilon}^{(l,1)},\mathbf{\Upsilon}^{(l,2)},\cdots,\mathbf{\Upsilon}^{(l,T_l)}$ and $\mathbf{\Theta}^{(l,1)},\mathbf{\Theta}^{(l,2)},\cdots,\mathbf{\Theta}^{(l,T_l)}$ are diagonal matrices. The calculation of $\mathbf{b}_1^{(l)},\mathbf{b}_2^{(l)},\cdots,\mathbf{b}_K^{(l)}$ in line 4 of Algorithm~\ref{algorithm_1} requires
\begin{align}\label{Complexity_2}
    c_2'=2KM
\end{align}
floating-point multiplications. Furthermore, the calculation of $\overline{\mathbf{b}}_k^{(l,t)}$ and $\overline{\mathbf{h}}_k^{(l,t)}$ in line 6 and 7 of Algorithm~\ref{algorithm_1} requires
\begin{align}\label{Complexity_3_t_1}
    c_{3,t1}'=\left(t-1\right)(N^2+2N)+\left(M+1\right)N
\end{align}
and
\begin{align}\label{Complexity_3_t_2}
    c_{3,t2}'=\left(T_l-t\right)(N^2+2N)
\end{align}
floating-point multiplications, respectively. The loop between line 8 of Algorithm~\ref{algorithm_1} has no floating-point multiplications. Thus, the loop between line 5 to 9 of Algorithm~\ref{algorithm_1} entails
\begin{align}\label{Complexity_3}
    \notag c_{3}'=&\sum\nolimits_{t=1}^{T_L}\left(c_{3,t1}'+c_{3,t2}'\right)\\
    =&\left(T_l^2-T_l\right)(N^2+2N)+T_l(M+1)N
\end{align}
floating-point multiplications. Therefore, according to (\ref{Complexity_1}), (\ref{Complexity_2}) and (\ref{Complexity_3}), the total number of floating-point multiplications in each iteration is
\begin{align}\label{Complexity_total}
    \notag c'=&c_1'+c_2'+c_3'\\
    \notag=&\left(T_l^2-1\right)N^2+\left(2T_l^2+T_lM+T_L+M\right)N\\
    &+\left(2K+1\right)M.
\end{align}
Hence for $N>M$ and $N>K$, the overall computational complexity of our proposed layer-by-layer iterative optimization of the hybrid beamformer at all APs is
\begin{align}
    \mathcal{O}\left(\tau\left(\sum_{l=1}^{L}T_l^2\right)N^2\right).
\end{align}
This shows that the proposed layer-by-layer iterative optimization algorithm conceived for the SIM-based hybrid beamformer is of polynomial time-complexity with respect to the number of APs, the number of SIM layers at each AP, and the number of reconfigurable elements in each SIM layer.

\subsubsection{Computational complexity of the CPU processing}
To recover the information $s_k$, the computational complexity at the CPU includes the calculation of the matrix $\mathbf{R}_k$ defined in (\ref{Beamforming_Design_12}), of the linear combination vector $\boldsymbol{\eta}_k$ defined in (\ref{Beamforming_Design_14}) and of the information recovery $\hat{s}_k$ defined in (\ref{Beamforming_CPU}). Specifically, the calculation of $\mathbf{R}_k$ requires
\begin{align}\label{Complexity_CPU_1}
    c_1''=\left(L^2+2L+3M\right)K
\end{align}
floating-point multiplication operations according to (\ref{Beamforming_Design_12}), (\ref{Beamforming_Design_12_1}), (\ref{Beamforming_Design_12_2}) and (\ref{Beamforming_Design_12_3}). The calculation of the linear combining vector $\boldsymbol{\eta}_k$ requires
\begin{align}\label{Complexity_CPU_2}
    c_2''=\frac{1}{3}\left(L^3-L\right)+L^2
\end{align}
floating-point multiplication operations by employing the Cholesky decomposition of the Hermitian positive-definite matrix $\mathbf{R}_k^{-1}$. Furthermore, the calculation of the information recovery $\hat{s}_k$ requires
\begin{align}\label{Complexity_CPU_3}
    c_3''=L
\end{align}
floating-point multiplication operations according to (\ref{Beamforming_CPU}). Therefore, according to (\ref{Complexity_CPU_1}), (\ref{Complexity_CPU_2}) and (\ref{Complexity_CPU_3}), the total number of floating-point multiplications required for recovering the information $s_k$ is
\begin{align}\label{Complexity_total}
    \notag c''=&c_1''+c_2''+c_3''\\
    =&\frac{1}{3}L^3+\left(K+1\right)L^2+\left(2K+\frac{2}{3}\right)L+3MK.
\end{align}
Therefore, the overall computational complexity of recovering the information $s_1,s_2,\cdots,s_K$ at the CPU is
\begin{align}
    \mathcal{O}\left(KL^3\right)+\mathcal{O}\left(K^2L^2\right)+\mathcal{O}\left(K^2M\right).
\end{align}
This shows that the CPU processing conceived for information recovery in the cell-free systems is of polynomial time-complexity with respect to the number of UEs, the number of APs and the number of RF chains at each AP.

\subsection{Convergence Analysis}
First, in line 3 and 4 of Algorithm~\ref{algorithm_1} convinced for the digital beamforming design, the received SINR $\gamma_k$ becomes non-decreasing after the digital beamformer $\mathbf{b}_1^{(l)},\mathbf{b}_2^{(l)},\cdots,\mathbf{b}_K^{(l)}$ has been optimized, given the holographic beamformer
$\mathbf{\Theta}^{(l,1)},\mathbf{\Theta}^{(l,2)},\cdots,\mathbf{\Theta}^{(l,T_l)}$, i.e.,
\begin{align}\label{Converge_1}
    \gamma_k(\ddot{\mathbf{b}}^{(l,i+1)},\ddot{\mathbf{\Theta}}^{(l,i)})\geq
    \gamma_k(\ddot{\mathbf{b}}^{(l,i)},\ddot{\mathbf{\Theta}}^{(l,i)}),
\end{align}
where $\gamma_k(\ddot{\mathbf{b}}^{(l,i_1)},\ddot{\mathbf{\Theta}}^{(l,i_2)})$ represents the received SINR based on the digital beamformer $\mathbf{b}_1^{(l)},\mathbf{b}_2^{(l)},\cdots,\mathbf{b}_K^{(l)}$ in the $i_1$th iteration and on the holographic beamformer
$\mathbf{\Theta}^{(l,1)},\mathbf{\Theta}^{(l,2)},\cdots,\mathbf{\Theta}^{(l,T_l)}$ in the $i_2$th iteration. Secondly, in line 5 to 9 of Algorithm~\ref{algorithm_1}, the received SINR $\gamma_k$ becomes non-decreasing after the holographic beamformer $\mathbf{\Theta}^{(l,1)},\mathbf{\Theta}^{(l,2)},\cdots,\mathbf{\Theta}^{(l,T_l)}$ has been optimized, given the digital beamformer
$\mathbf{b}_1^{(l)},\mathbf{b}_2^{(l)},\cdots,\mathbf{b}_K^{(l)}$, i.e.,
\begin{align}\label{Converge_2}
    \gamma_k(\ddot{\mathbf{b}}^{(l,i)},\ddot{\mathbf{\Theta}}^{(l,i+1)})\geq
    \gamma_k(\ddot{\mathbf{b}}^{(l,i)},\ddot{\mathbf{\Theta}}^{(l,i)}).
\end{align}
Therefore, (\ref{Converge_1}) and (\ref{Converge_2}) imply that in each iteration of the proposed layer-by-layer iterative optimization algorithm at each AP, the objective function value of the received SINR $\gamma_k$ is non-decreasing. Additionally, the objective function value sequence obtained throughout the iteration process is monotonic and it is also bounded, hence the overall algorithm is guaranteed to converge.

\begin{table}[!t]
\footnotesize
\begin{center}
\caption{Simulation Parameters.}\label{Table_Simulation}
\begin{tabular}{|p{102pt}|p{116pt}|}
\hline
    \textbf{Parameters} & \textbf{Values} \\
\hline
    Carrier frequency & $f_c=30$ GHz\\
\hdashline
    Number of APs & $L=16$ \\
\hdashline
    Number of UEs & $K=8$ \\
\hdashline
    Number of AP antennas & $M=4\times4$ \\
\hdashline
    Number of SIM elements in each layer & $N=16\times16$ \\
\hdashline
    Number of SIM layers & $T_l=4$ ($l=1,2,\cdots L$) \\
\hdashline
    Inter-layer distance & $z^{(l,0)}=z^{(l,1)}=\cdots=z^{(l,T_l-1)}=\lambda$ ($l=1,2,\cdots L$) \\
\hdashline
    Directional radiation gain & $\kappa=10\text{dB}$ \\
\hdashline
    Distance between AP antennas  & $d_x=d_y=\frac{\lambda}{2}$ \\
\hdashline
    Size of the SIM elements  & $\delta_x=\delta_y=\frac{\lambda}{4}$ \\
\hdashline
    Noise power  & $\sigma_{w^{(l)}}^2=-80\text{dBm}$ ($l=1,2,\cdots L$) \\
\hdashline
    Hardware quality factor  & $\varepsilon_{u_k}=\varepsilon_{\mathbf{v}^{(l)}}=\varepsilon=1$ ($k=1,2,\cdots,K$, and $l=1,2,\cdots,L$) \\
\hdashline
    Transmit power at UEs  & $\rho=\rho_1=\rho_2=\cdots=\rho_K$ \\
\hdashline
    Path loss exponent & $\beta=3.5$ \\
\hdashline
    Path loss at the reference distance of 1 meter & $\mathrm{C}_0=10^{-3}$ \\
\hdashline
    Number of clusters & $\zeta_{c}=4$ ($c=1,2,\cdots,\zeta_{c}$) \\
\hdashline
    Number of paths in each cluster & $\zeta_{p}=8$ ($p=1,2,\cdots,\zeta_{p}$) \\
\hdashline
    Mean values of the elevation angles & $\mu_{\psi_{c}^k}$ are randomly distributed in $[0^\circ,180^\circ]$ ($c=1,2,\cdots,\zeta_{c}$, and $k=1,2,\cdots,K$) \\
\hdashline
    Mean values of the azimuth angles & $\mu_{\varphi_{c}^k}$ are randomly distributed in $[0^\circ,360^\circ]$ ($c=1,2,\cdots,\zeta_{c}$, and $k=1,2,\cdots,K$) \\
\hdashline
    Elevation angle spreads & $\sigma_{\psi_{c}^k}=7.5^\circ$ ($c=1,2,\cdots,\zeta_{c}$, and $k=1,2,\cdots,K$) \\
\hdashline
    Azimuth angle spreads & $\sigma_{\varphi_{c}^k}=7.5^\circ$ ($c=1,2,\cdots,\zeta_{c}$, and $k=1,2,\cdots,K$) \\
\hdashline
    SIM power radiation coefficients &  $\upsilon_n^{(l,t)}=1$ ($n=1,2,\cdots,N$, $t=1,2,\cdots,T_l$, and $l=1,2,\cdots,L$) \\
\hline
\end{tabular}
\end{center}
\end{table}

\section{Numerical and Simulation Results}\label{Numerical_and_Simulation_Results}
In this section, the average achievable rate of the SIM-based holographic MIMO of cell-free networks is quantified. A total of $L$ APs are uniformly distributed in the area $\mathcal{S}$ with the Cartesian coordinate of $\{(x,y):-100\text{m}\leq x\leq100\text{m},-100\text{m}\leq y\leq100\text{m}\}$. Furthermore, $K$ UEs are uniform-randomly distributed in the area $\mathcal{S}$. We employ a distance-dependent path-loss model for the UE-AP channel, given by $\varrho_k^{(l)}=\min\{\mathrm{C}_0,\mathrm{C}_0(d_k^{(l)})^{-\beta}\}$, where $d_k^{(l)}$ is the length of the link spanning from AP-$l$ to UE-$k$, $\beta$ is the path loss exponent, and $\mathrm{C}_0$ is the path loss at the reference distance of 1 meter~\cite{haenggi2012stochastic}. Referring to~\cite{an2024stacked_earlyaccess}, \cite{deng2022reconfigurable_twc}, the simulation parameters are those given in Table~\ref{Table_Simulation}, unless otherwise specified.

\begin{figure}[!t]
    \centering
    \includegraphics[width=2.45in]{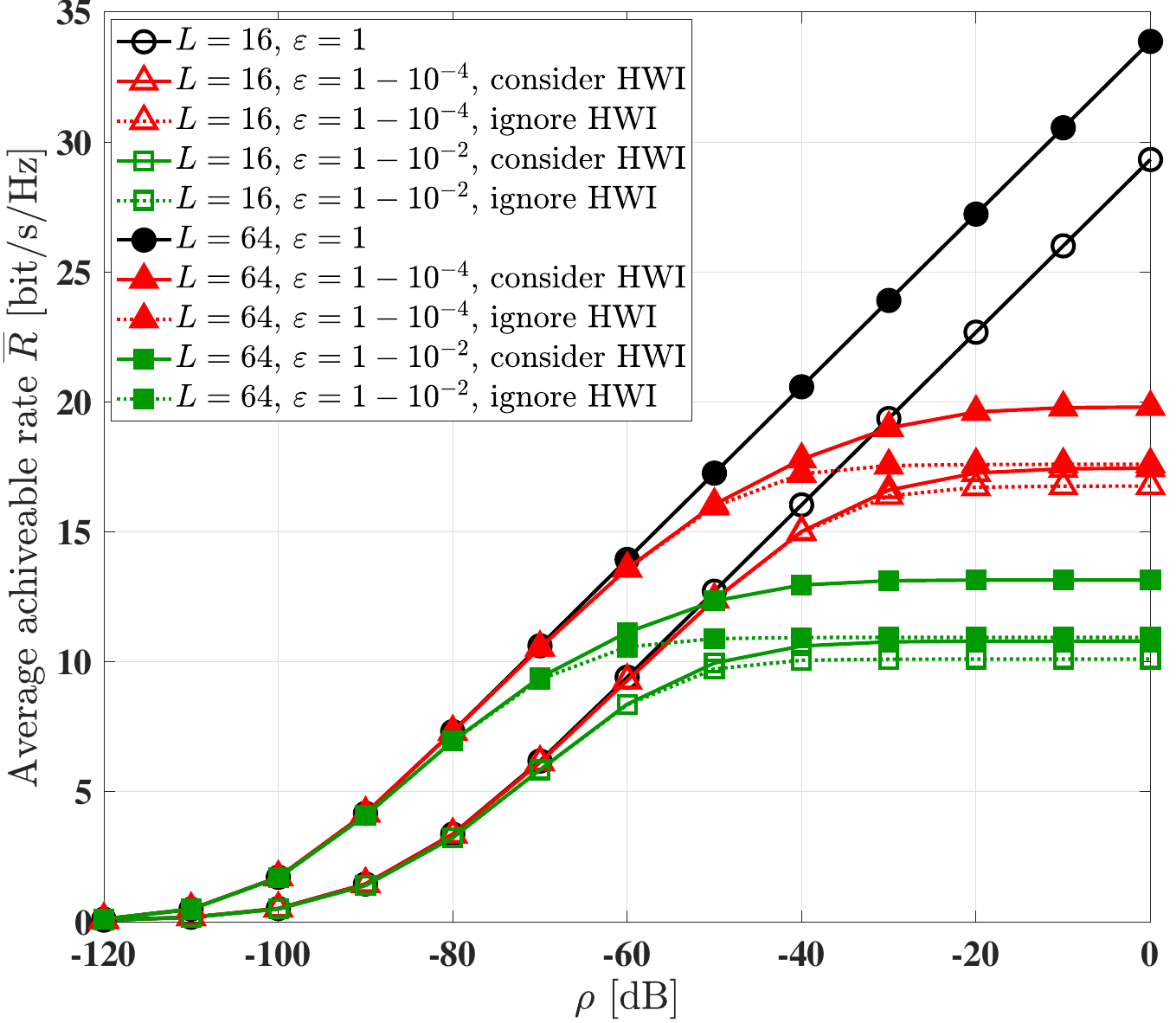}
    \caption{Comparison of the average achievable rate $\overline{R}$ versus the transmit power $\rho$ with different number of APs.}\label{Simu_Fig_1}
\end{figure}

\begin{figure}[!t]
    \centering
    \includegraphics[width=2.45in]{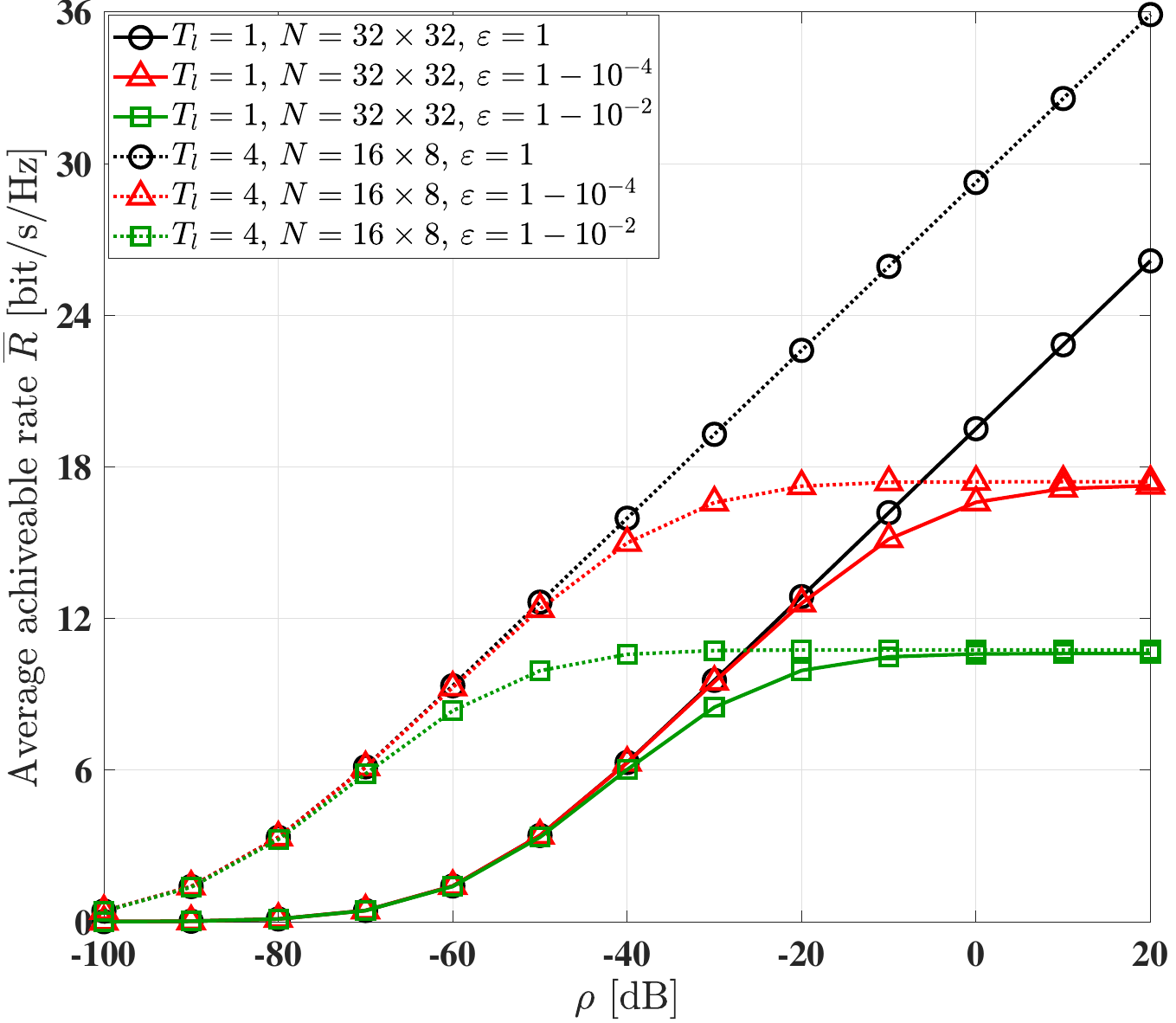}
    \caption{Comparison of the average achievable rate $\overline{R}$ versus the transmit power $\rho$ with different number of SIM layers.}\label{Simu_Fig_2}
\end{figure}

\begin{figure}[!t]
    \centering
    \subfloat[Hardware quality factor $\varepsilon=1$.]
    {\begin{minipage}{1\linewidth}
        \centering
        \includegraphics[width=2.45in]{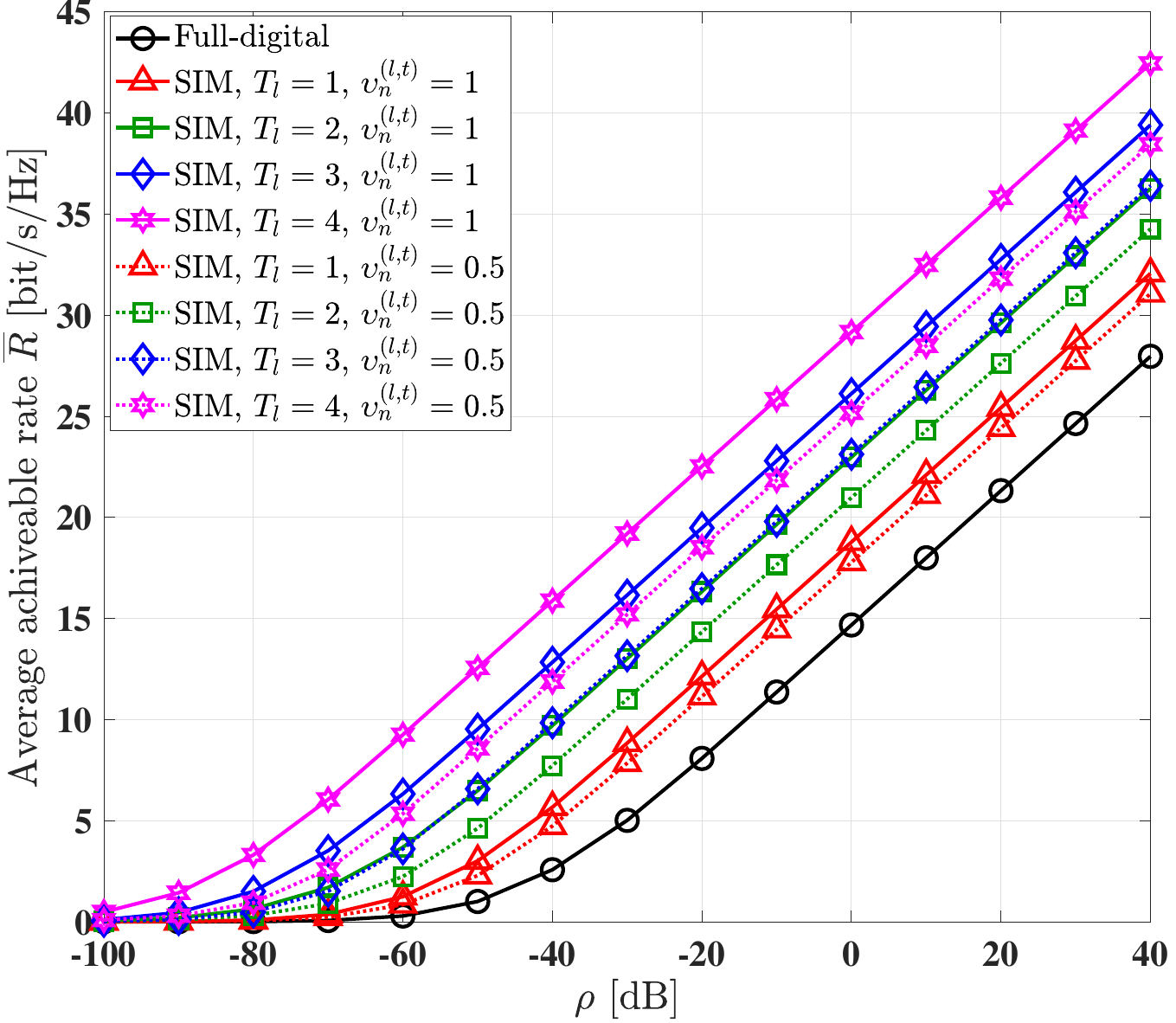}
    \end{minipage}}\\
    \subfloat[Hardware quality factor $\varepsilon=1-10^{-4}$.]
    {\begin{minipage}{1\linewidth}
        \centering
        \includegraphics[width=2.45in]{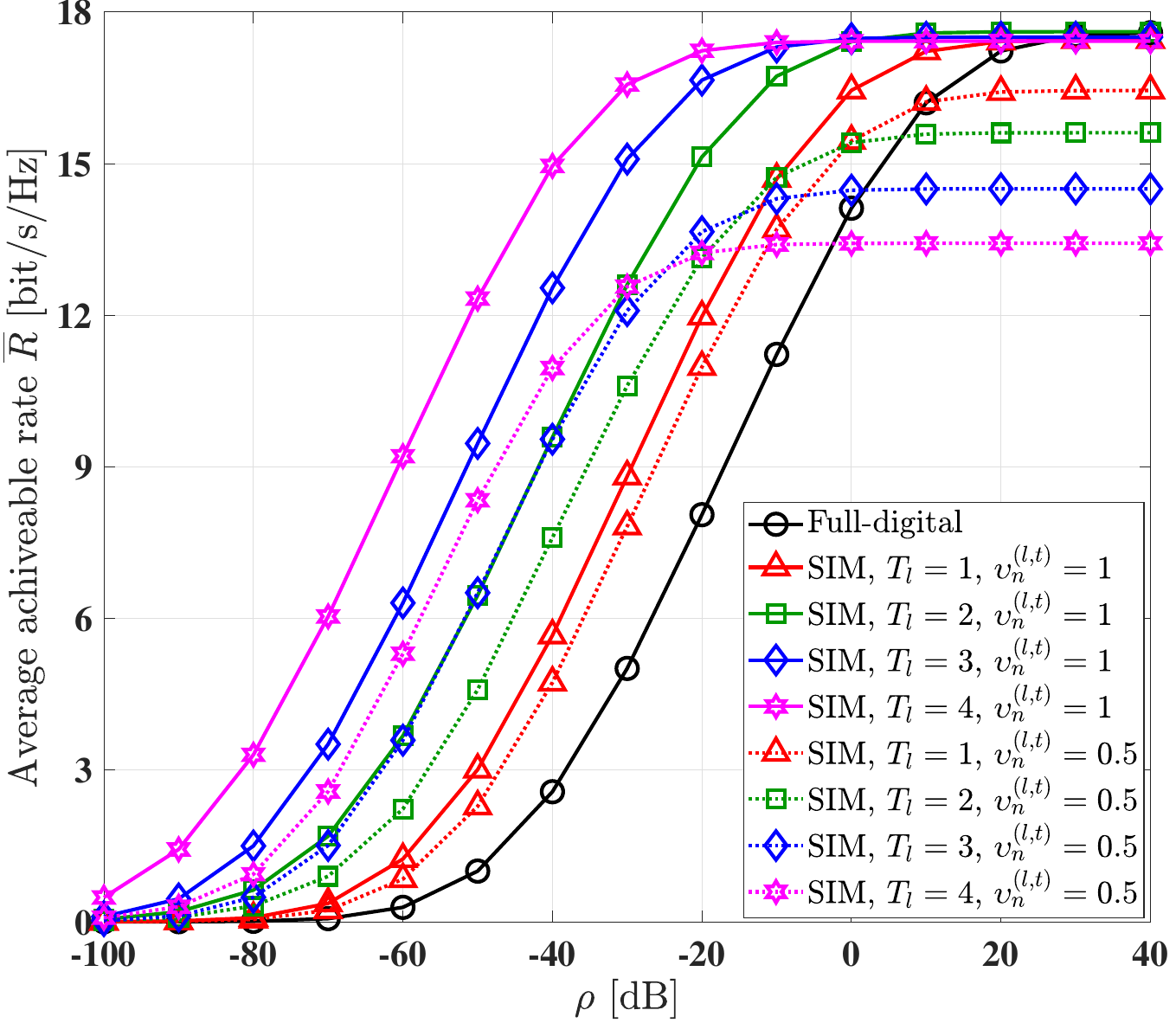}
    \end{minipage}}\\
    \subfloat[Hardware quality factor $\varepsilon=1-10^{-2}$.]
    {\begin{minipage}{1\linewidth}
        \centering
        \includegraphics[width=2.45in]{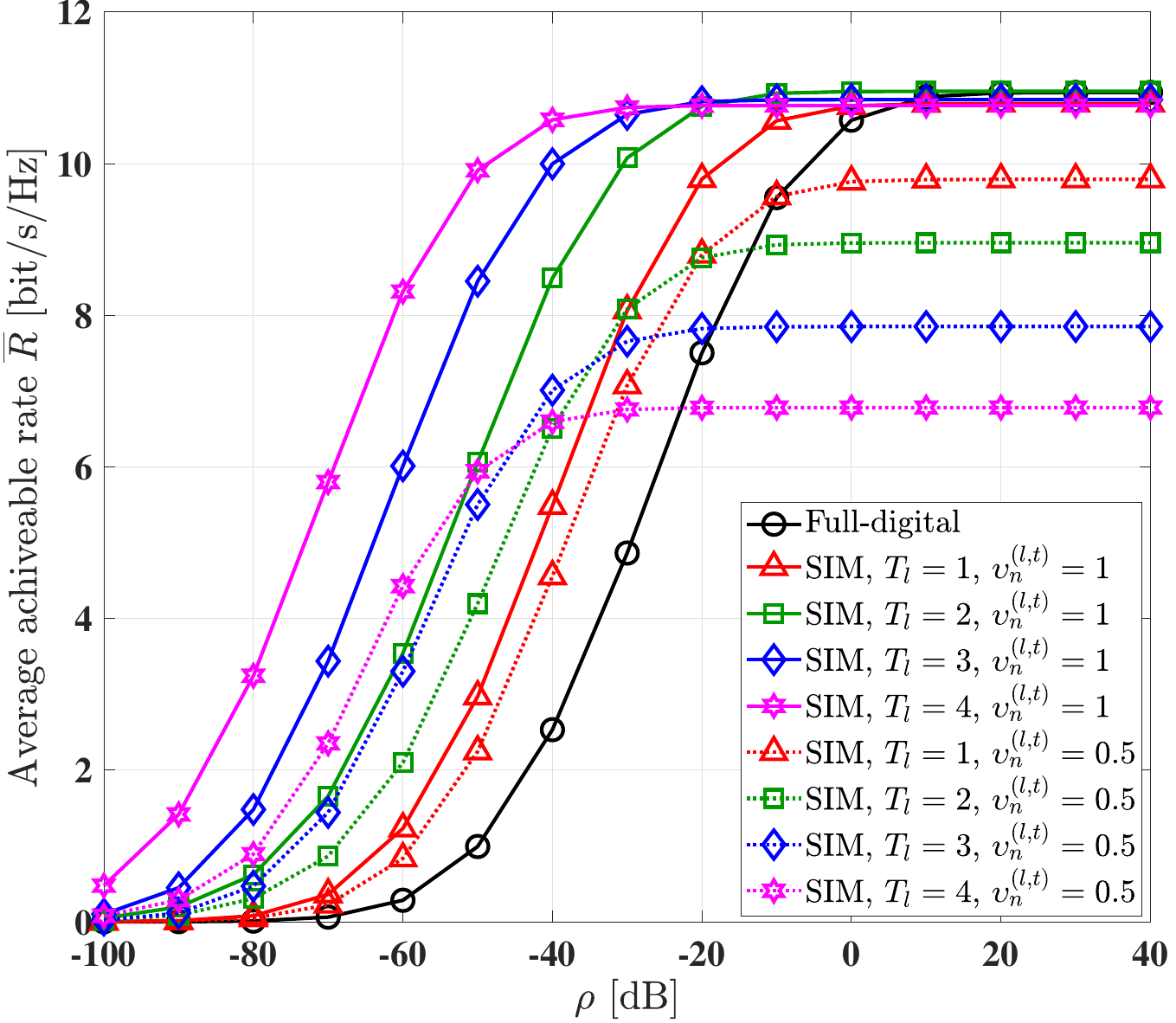}
    \end{minipage}}
    \caption{Average achievable rate $\overline{R}$ versus the transmit power $\rho$ in the conventional full-digital beamformer and the SIM-based hybrid beamformer.}\label{Simu_Fig_345}
\end{figure}

Fig.~\ref{Simu_Fig_1} compares the average achievable rate $\overline{R}=\frac{1}{K}\sum_{k=1}^KR_k=\frac{1}{K}\sum_{k=1}^K\log_2(1+\gamma_k)$ versus the transmit power $\rho$ for different number of APs in cell-free networks, with the number of AP $L=16$ and $L=64$, respectively. Observe from Fig.~\ref{Simu_Fig_1} that the average achievable rate can be improved by increasing the number of APs when the average distance between the APs and UEs can be reduced. When the hardware is non-ideal, i.e. $\varepsilon<1$, the average achievable rate tends to a constant value upon increasing of the transmit power $\rho$. Furthermore, as seen in Fig.~\ref{Simu_Fig_1}, an obvious performance degradation exists when the HWIs are ignored during information recovery, especially with the increase of the number of APs. Fig.~\ref{Simu_Fig_2} compares the average achievable rate $\overline{R}$ versus the transmit power $\rho$ for different number of SIM layers in each AP, where the number $N$ of elements in each SIM layer is set to $N=16\times16$, $N=8\times8$ and $N=4\times4$ when the number of SIM layers $T_l$ are 1, 2 and 4 respectively for ensuring that the same total number of SIM elements are employed. Observe in Fig.~\ref{Simu_Fig_2} that the average achievable rate is improved upon increasing the number of SIM layers. It inspires our future research on the SIM design to determine how many SIM layers is optimal in terms of maximizing the performance gain given the total number of SIM elements.

To provide further insights, Fig.~\ref{Simu_Fig_345} characterizes the average achievable rate of both our SIM-based hybrid beamforming architecture and of the conventional cell-free full-digital beamforming architecture, for the hardware quality factors of $\varepsilon=1$, $1-10^{-4}$ and $1-10^{-2}$, respectively. In terms of the perfect hardware quality factor, i.e., $\varepsilon=1$, Fig.~\ref{Simu_Fig_345} (a) shows that the SIM-based hybrid beamformer has higher average achievable rate than the conventional full-digital beamformer, even when the SIM has the signal radiation attenuation associated with the power radiation coefficients of $\upsilon_n^{(l,t)}=0.5$. This is a benefit of the beamforming gain attained by the configuration of the SIM elements. Furthermore, the average achievable rate can be improved upon increasing the number of SIM layers. When the hardware quality is imperfect, i.e., $\varepsilon<1$, it is illustrated in Fig.~\ref{Simu_Fig_345} (b) and Fig.~\ref{Simu_Fig_345} (c) that the SIM-based hybrid beamformer outperforms the full-digital beamformer in the low-SNR region, while the achievable rate in the high-SNR region is limited by the hardware quality. Moreover, in the high-SNR region, increasing the number of SIM layers even degrades the achievable rate when the SIM power radiation coefficients $\upsilon_n^{(l,t)}<1$, since the signals are attenuated in each layer of the SIM and it cannot be compensated by the configuration of SIM elements due to the limitation of the hardware quality. It can be observed that although the SIM can increase the equivalent channel gain, but it cannot compensate for the deleterious effects of HWIs in the high-SNR region.

\begin{figure}[!t]
    \centering
    \includegraphics[width=2.45in]{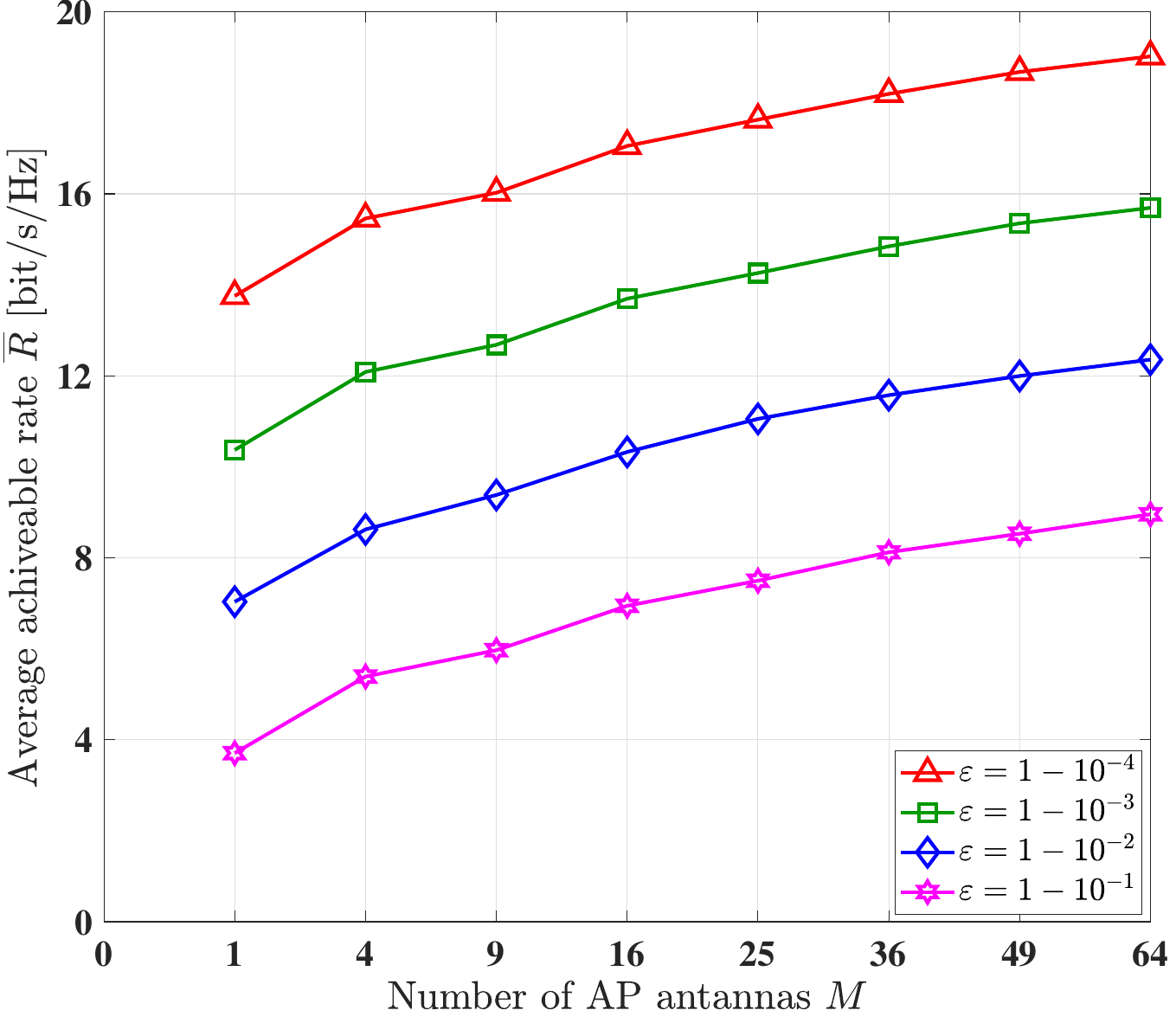}
    \caption{Average achievable rate $\overline{R}$ versus the number of AP antennas $M$.}\label{Simu_Fig_6}
\end{figure}

The performance comparison of the average achievable rate $\overline{R}$ versus the number of antennas $M$ at each AP is presented in Fig.~\ref{Simu_Fig_6}, for different hardware quality factors $\varepsilon$. It shows that the average achievable rate can be improved upon increasing the number of antennas at each AP, but only at the cost of higher energy consumption. However, the signal distortion resulting from the imperfect hardware quality can be compensated by employing more AP antennas.

\begin{figure}[!t]
    \centering
    \includegraphics[width=2.45in]{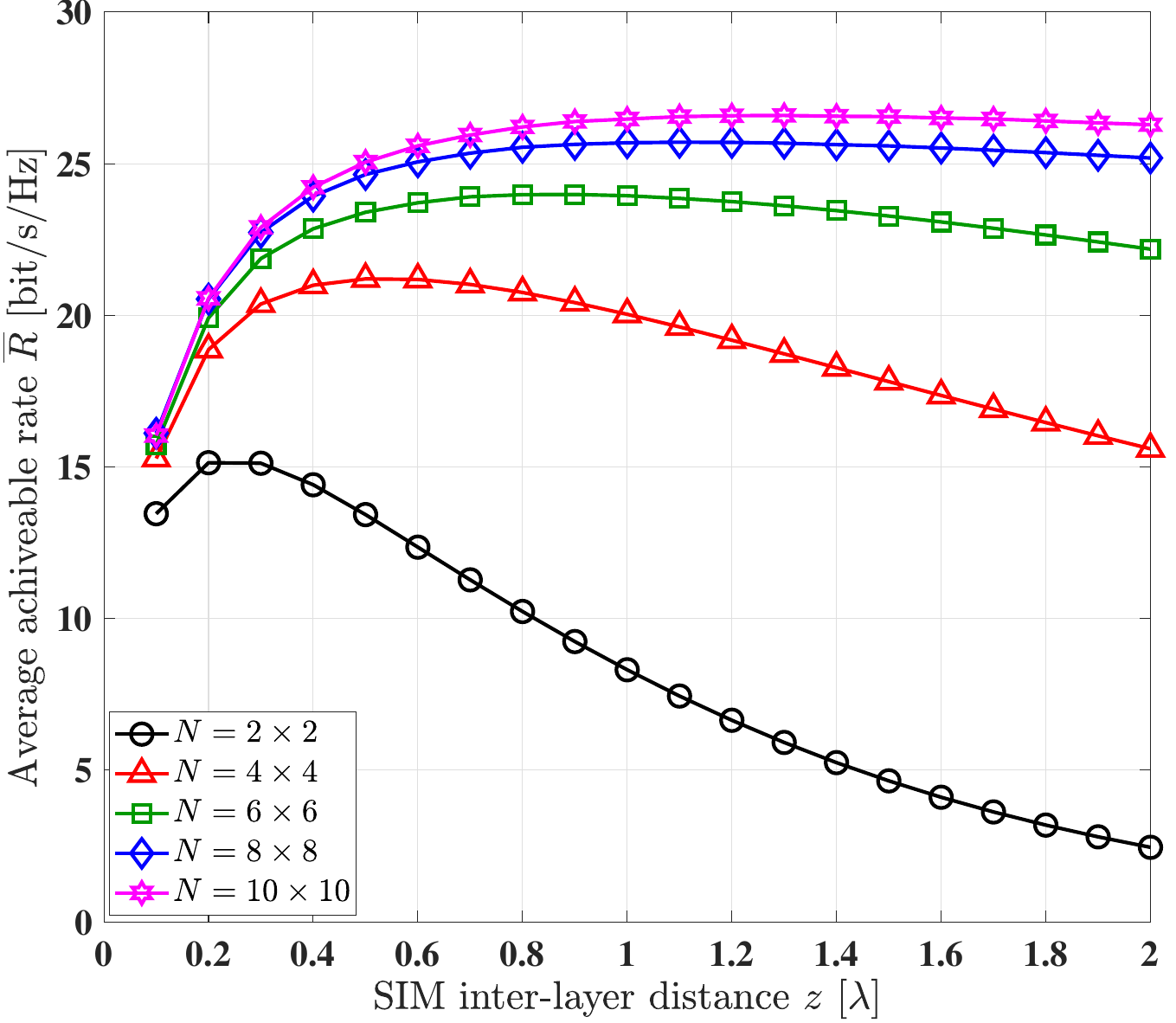}
    \caption{Average achievable rate $\overline{R}$ versus the SIM inter-layer distance $z$.}\label{Simu_Fig_7}
\end{figure}

Fig.~\ref{Simu_Fig_7} portrays the average achievable rate versus the inter-layer distance, where we have $z=z^{(l,0)}=z^{(l,1)}=\cdots=z^{(l,T_l-1)}$ for $l=1,2,\cdots,L$. The figure shows that there exists an optimal inter-layer distance in terms of the highest average achievable rate. Furthermore, it also shows that the optimal inter-layer distance increases upon employing more intelligent surface elements. Observe that when the number of intelligent surface elements is small, the beamforming gain attained by the SIM is predominantly limited by the path loss between layers, and thus a higher average achievable rate can be attained by reducing the inter-layer distance. By contrast, upon increasing the number of intelligent surface elements, the inter-layer path loss effect becomes negligible and increasing the inter-layer distance improves the signal radiation between adjacent layers, increasing the achievable rate.

\begin{figure}[!t]
    \centering
    \includegraphics[width=2.45in]{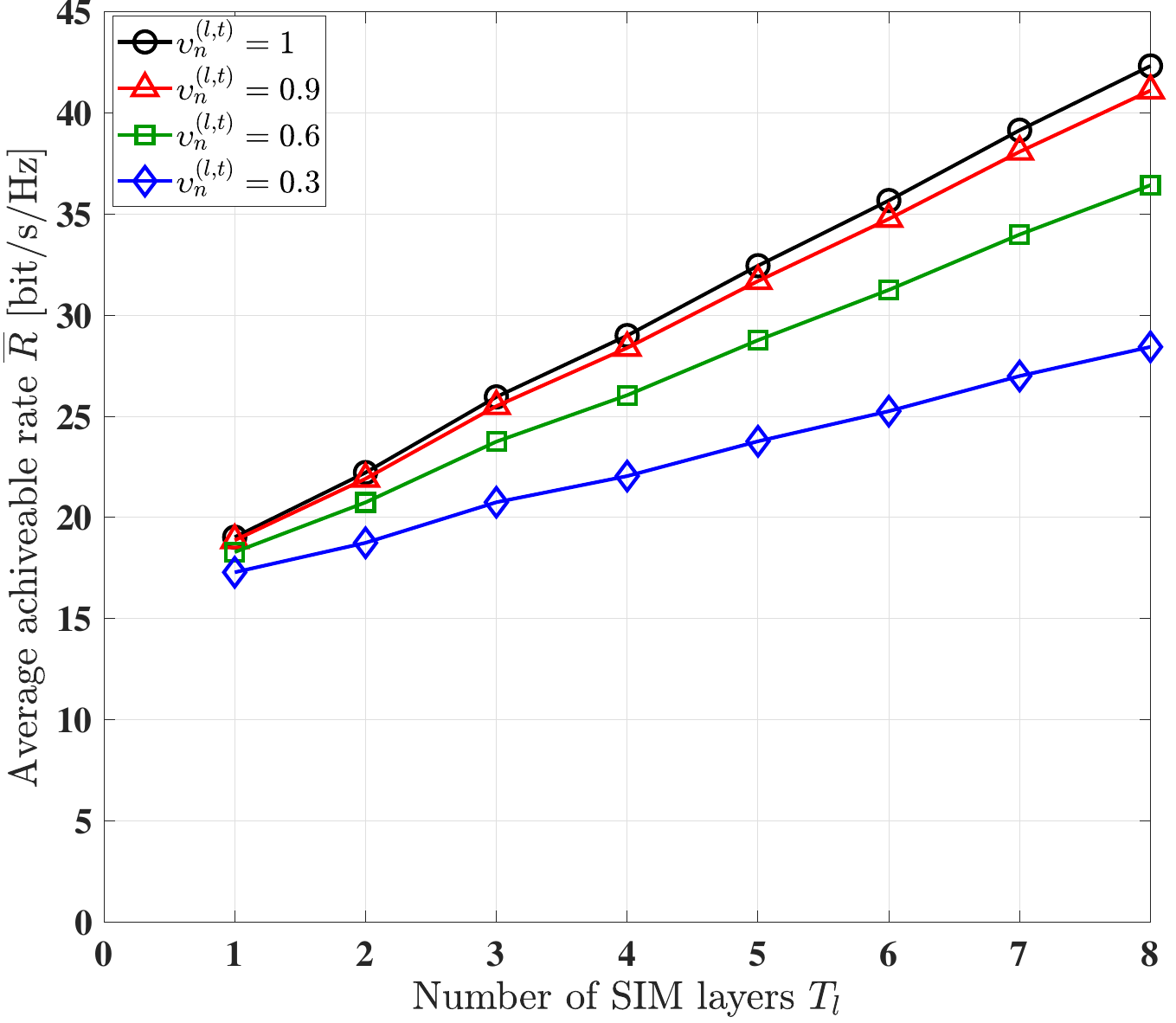}
    \caption{Average achievable rate $\overline{R}$ versus the number of SIM layers $T_l$.}\label{Simu_Fig_8}
\end{figure}

To investigate the impact of the signal attenuation caused by the signal travelling through each layer of the SIM on the achievable rate performance, Fig.~\ref{Simu_Fig_8} presents the average achievable rate $\overline{R}$ versus the number of SIM layers at each AP, characterize by different power radiation coefficients $\upsilon_n^{(l,t)}$. It shows that although the average achievable rate degrades with the reduction of the radiation coefficients $\upsilon_n^{(l,t)}$, this effect can be compensated by increasing the number of SIM layers.

\begin{figure}[!t]
    \centering
    \includegraphics[width=2.45in]{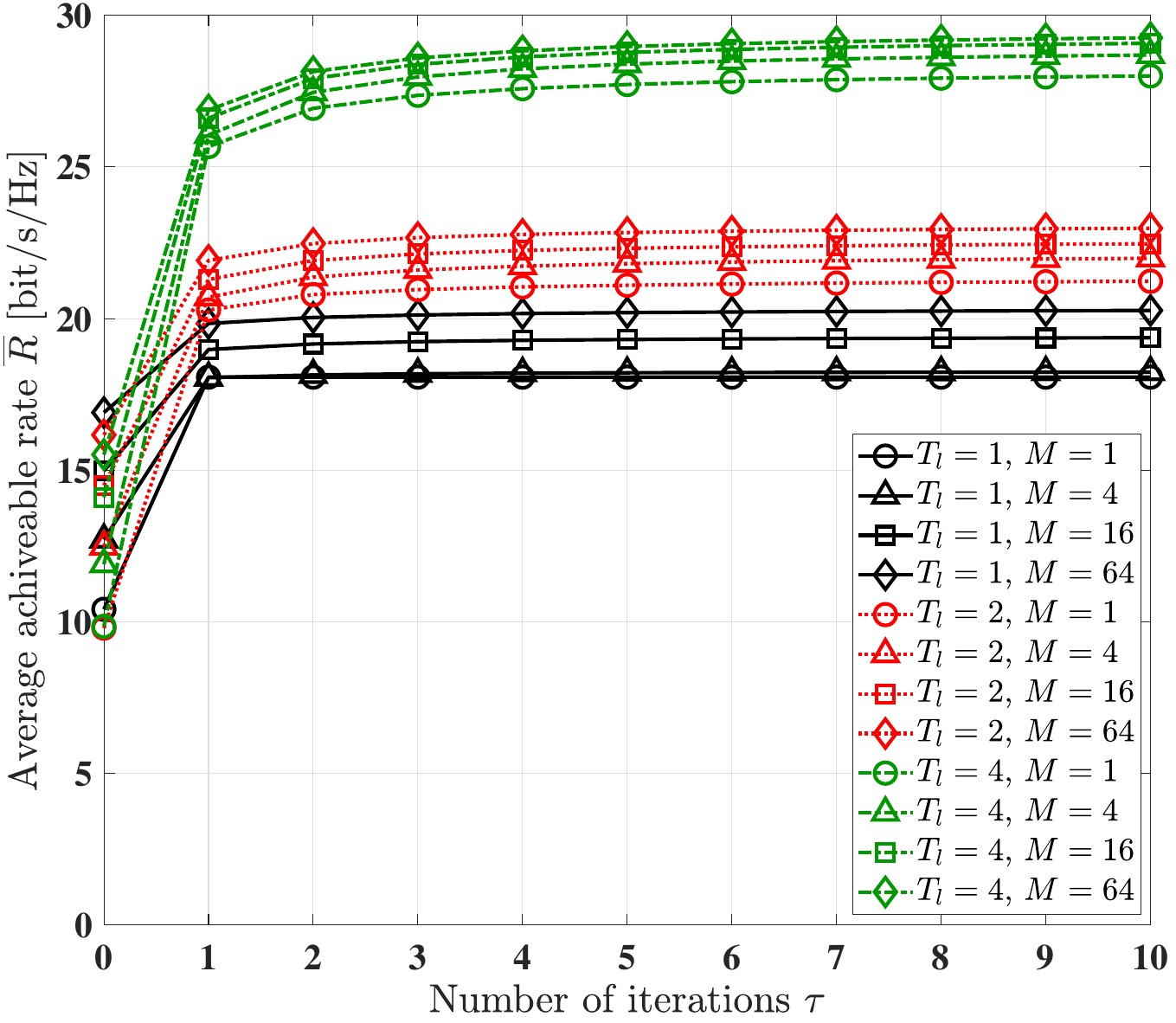}
    \caption{Average achievable rate $\overline{R}$ versus the number of iterations $\tau$.}\label{Simu_Fig_9}
\end{figure}

In Fig.~\ref{Simu_Fig_9} we portray the average achievable rate $\overline{R}$ versus the number of iterations $\tau$ attained by the proposed layer-by-layer iterative optimization algorithm used by the hybrid beamformer. Observe that although the convergence speed is reduced as the number of SIM layers increases, it achieves convergence within 10 iterations.

\begin{figure}[!t]
    \centering
    \includegraphics[width=2.45in]{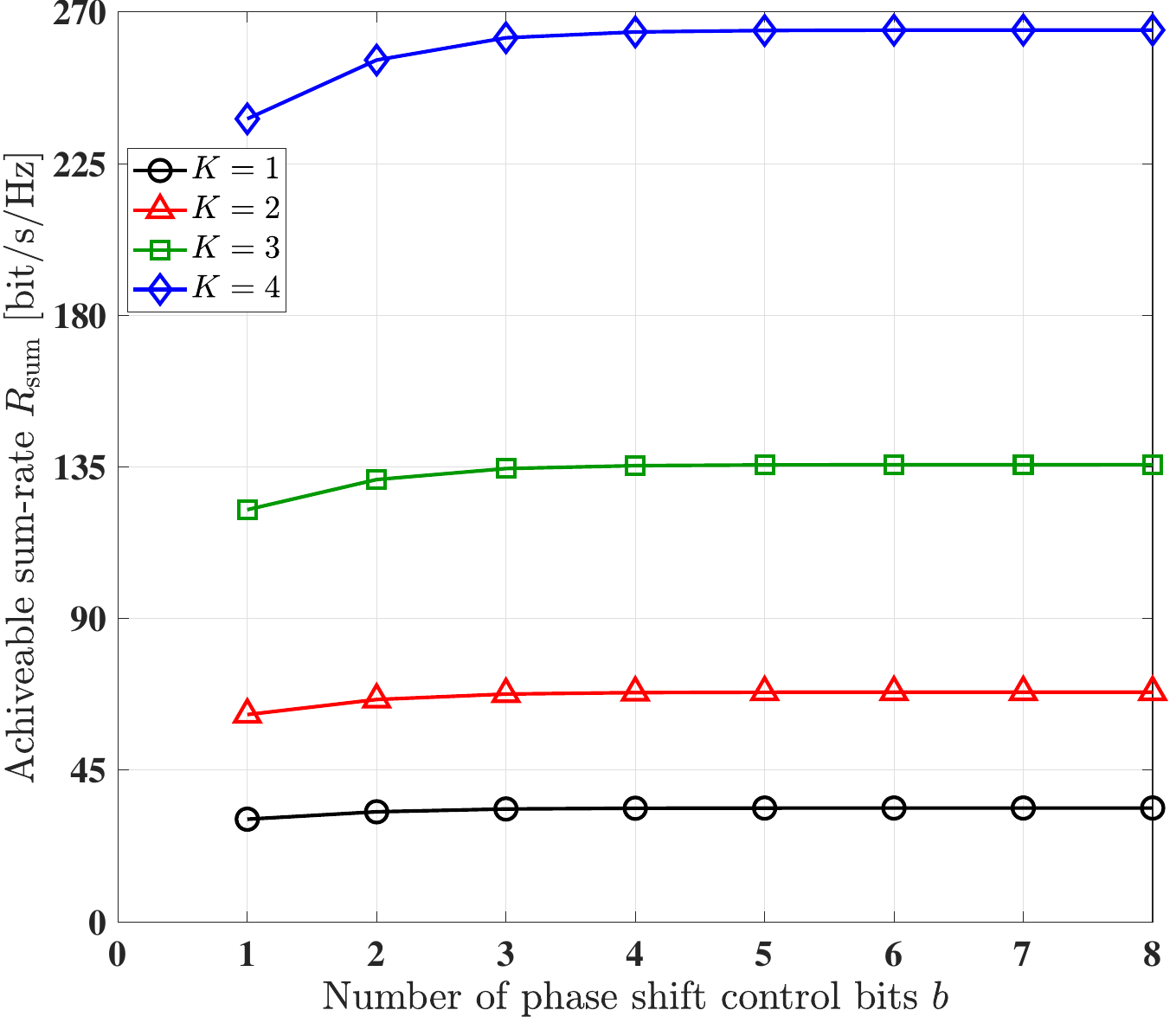}
    \caption{Achievable sum-rate $R_\mathrm{sum}$ versus the number of phase shift control bit $b$.}\label{Simu_Fig_10}
\end{figure}

In the above simulations, we assume that the reconfigurable SIM elements have continuous phase shifts in the range of $[0,2\pi)$. However, in practical hardware implementations, the phase shift of each reconfigurable element is limited to a finite number of discrete values. For simplicity, we assume that the discrete phase shift set is obtained by uniformly quantizing the interval $[0,2\pi)$ into $2^b$ levels, with $b$ respecting the number of phase shift control bits. Thus, we have $\theta_n^{(l,t)}\in\{2\pi\cdot\frac{0}{2^B},2\pi\cdot\frac{1}{2^B},\cdots,
2\pi\cdot\frac{2^b-1}{2^b}\}$. Fig.~\ref{Simu_Fig_10} characterizes the achievable sum-rate, denoted as $R_{\mathrm{sum}}=\sum_{k=1}^KR_k=\sum_{k=1}^K\log_2(1+\gamma_k)$  versus the number of phase shift control bits $b$, with different number of UEs $k$. It shows that 4-bit phase shift quantization can approach the rate of the infinite phase shift resolution.

\section{Conclusions}\label{Conclusion}
We conceived an uplink SIM-based cell-free HMIMO architecture, where the distributed signal processing was employed. Each AP carries out a local detection of the UE information, where the hybrid beamformer and the RC vectors of each distributed AP are optimized based on our low-complexity layer-by-layer iterative optimization algorithm. The CPU recovers the final UE information by fusing the local detections gleaned from all APs, where the RC weight vector used for combining the local detections is designed based on the MMSE criterion, taking into account the HWIs of the RF chains at the APs and of the UEs. The simulation results showed that the achievable rate of the SIM-based cell-free HMIMO network improves upon increasing the number of SIM layers as well as the number of elements in each layer. Furthermore, due to the HWI of the RF chains at the APs and the UEs, the achievable rate saturates in the high-SNR region.

\bibliographystyle{IEEEtran}
\bibliography{IEEEabrv,TAMS}

\end{document}